\documentclass[%
twocolumn,
superscriptaddress, 
reprint,
amsmath,amssymb,
prx,
longbibliography,
floatfix,
]{revtex4-2}

\usepackage{graphicx}
\usepackage{dcolumn}
\usepackage{bm}
\usepackage{enumerate}
\usepackage{grffile}
\usepackage{color}
\usepackage{esint}
\usepackage{textpos}
\usepackage{subfigure}
\usepackage{flushend}
\usepackage{tabularx}
\usepackage{float}
\usepackage{setspace}
\usepackage{mathrsfs}  
\usepackage{soul}

\graphicspath{{Figures/}}

\renewcommand{\Vec}[1]{\boldsymbol{#1}}
\renewcommand{\vec}[1]{\boldsymbol{#1}}

\allowdisplaybreaks

\begin{document}

\title{Coupled Cluster Theory for Molecular Polaritons: Changing Ground and Excited States}

\author{Tor S. Haugland}
\affiliation{Department of Chemistry, Norwegian University of Science and Technology, 7491 Trondheim, Norway}
\author{Enrico Ronca}%
\affiliation{Istituto per i Processi Chimico Fisici del CNR (IPCF-CNR), Via G. Moruzzi, 1, 56124, Pisa, Italy}
\affiliation{%
Max Planck Institute for the Structure and Dynamics of Matter and Center Free-Electron Laser Science, Luruper Chaussee 149, 22761 Hamburg, Germany}%
\author{Eirik F. Kj\o nstad}
\affiliation{Department of Chemistry, Norwegian University of Science and Technology, 7491 Trondheim, Norway}
\author{Angel Rubio}
\affiliation{%
Max Planck Institute for the Structure and Dynamics of Matter and Center Free-Electron Laser Science, Luruper Chaussee 149, 22761 Hamburg, Germany}%
\affiliation{%
Center for Computational Quantum Physics (CCQ), The Flatiron Institute, 162 Fifth avenue, New York NY 10010.}%
\affiliation{%
Nano-Bio Spectroscopy Group, Departimento de F{\'i}sica de Materiales, Universidad del Pa{\'i}s Vasco, 20018 San Sebastian, Spain}%
\author{Henrik Koch}
\email{henrik.koch@sns.it}
\affiliation{Scuola Normale Superiore, Piazza dei Cavalieri, 7, 56124 Pisa, Italy}
\affiliation{Department of Chemistry, Norwegian University of Science and Technology, 7491 Trondheim, Norway}

\date{\today}

\begin{abstract}
We present an \textit{ab initio} correlated approach to study molecules that interact strongly with quantum fields in an optical cavity. 
Quantum electrodynamics coupled cluster theory provides a non-perturbative description of cavity-induced effects in ground and excited states. 
Using this theory, we show how quantum fields can be used to manipulate charge transfer and photochemical properties of molecules.
We propose a strategy to lift electronic degeneracies and induce modifications in the ground state potential energy surface close to a conical intersection. 
\end{abstract}

                              
\maketitle

\section{\label{sec:introduction} Introduction}

Nowadays, manipulation by strong electron-photon coupling~\cite{RuggenthalerNatureRevChem2018}, and laser fields~\cite{Kim2012, Kim2015,Hubener2017}, is becoming a popular technique to design and explore new states of matter.
Recent advances in experimental and theoretical research include new ways to generate exciton-polariton condensates~\cite{ByrnesNaturePhys2014}, induce phase transitions~\cite{WangPhysRevB2019,BasovNatureMater2017,LindnerNaturePhys2011}, tune exciton energies in monolayers of 2D materials and  
interfaces~\cite{LatiniNanoLett2019,ByrnesNaturePhys2014,SunNaturePhoton2017}, and even enhance the electron-phonon coupling with possible effects on superconductivity~\cite{SentefScienceAdv2018,CavalleriContempPhys2018}.
Nevertheless, general techniques for manipulating molecules via strong coupling 
have not yet reached maturity.

\begin{figure*}[ht!]
    \centering
    \includegraphics[trim={2cm 0.5cm 1.5cm 0.5cm},clip,width=8.6cm]{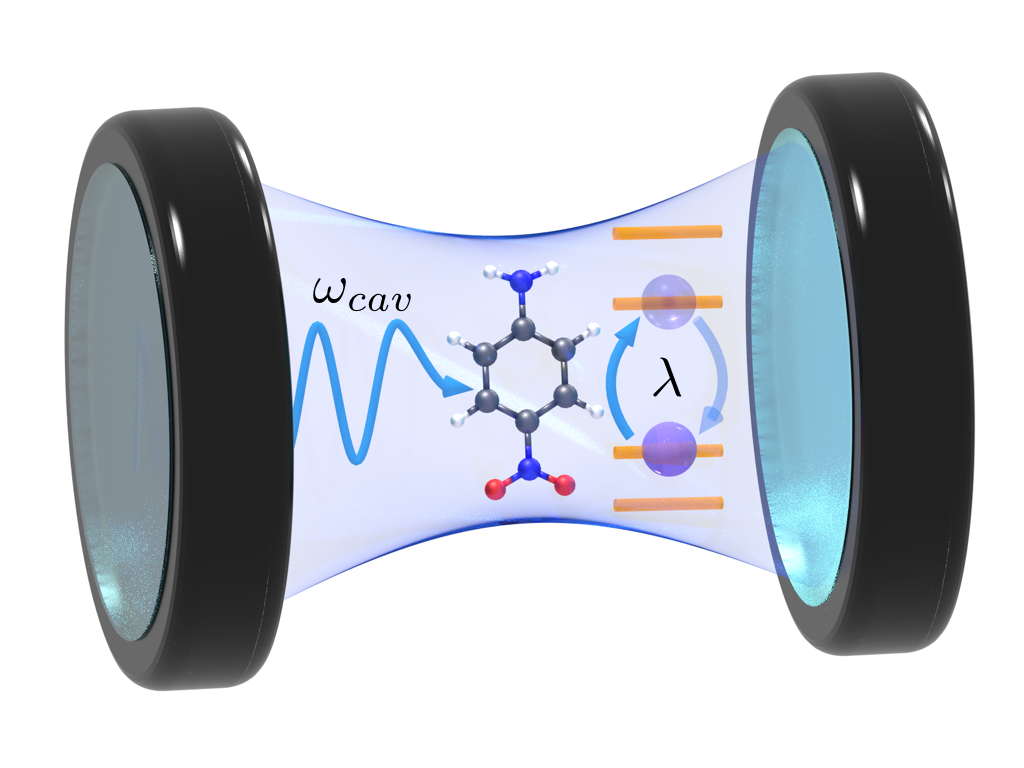}
    \caption{Illustration of an optical cavity interacting with a molecule.}
    \label{fig:cartoon}
\end{figure*}

Chemistry is one of the fields that has witnessed most progress in strong light-matter coupling applications.
In particular, Ebbesen and coworkers 
have found
that strong coupling to vibrational 
excited states in molecules
can inhibit ~\cite{ThomasAngewChemIntEd2016,EbbesenAccChemRes2016,GalegoPhysRevX2019}, catalyze~\cite{HiuraChemrxiv2018,ClimentChemrxiv2019}, and induce selective change in the reactive path of a chemical reaction~\cite{ThomasScience2019}. 
These
experiments use
an optical cavity, the simplest device where entanglement between matter and light can be observed.
In an optical cavity
(see Fig.~\ref{fig:cartoon}), the quantized electromagnetic field interacts 
with the molecular system,
producing new hybrid light-matter states called {\it polaritons}~\cite{BasovScience1992}. 
These states
exhibit new and interesting properties, leading to unexpected phenomena. 
In the past few years, 
improvements of optical cavities~\cite{SigleACSNano2015,ChikkaraddyNature2016,GalfskyPNAS2017,BenzNanoLett2015} 
have resulted in
devices 
that 
can reach the strong and ultra-strong coupling limits. 
This has made
polaritonic states accessible at room temperature, also for a  small number of molecules~\cite{ChikkaraddyNature2016,KongsuwanACSPhoton2018}.  

Theoretical modeling is an essential tool to provide fundamental understanding and outline new strategies for  applications 
in polaritonic chemistry.
The challenge is to develop an accurate 
theoretical description of entangled light-matter systems. 
In the quantum optics community, several groups 
have developed model Hamiltonians to reproduce the main features of polaritonic physics~\cite{GrynbergBook2010, WaltherRepProgPhys2006, RibeiroChemSci2018,AnguloArXiv2019,GalegoPhysRevX2015,GalegoPhysRevX2019}.
The objective of the current work is to formulate and implement a quantitative \emph{ab initio} method for polaritonic chemistry.

Presently, the only available \emph{ab initio} theory 
is quantum electrodynamical density functional theory (QEDFT)~\cite{RuggenthalerPhysRevA2014, FlickPNAS2017,RuggenthalerNatureRevChem2018,TokatlyPhysRevLett2013}, which can describe interacting electrons and photons, on an equal footing. This method is a natural extension of density functional theory (DFT)~\cite{ParrBook1989} to  quantum electrodynamics (QED). In QEDFT, the Kohn-Sham formalism treats electrons and photons as independent particles interacting through an exchange correlation potential.
The QEDFT method is computationally cheap and reproduces the main polaritonic features for large systems, even at the mean-field level. However, its accuracy is limited due to the unknown form of the exchange correlation functional for the electron-photon interaction.
In particular, in a mean-field treatment, no explicit electron-photon correlation is accounted for in the ground state.
The problem can be overcome with a properly designed exchange-correlation functional, but current functionals are still not sufficiently accurate. 
Recently, 
Rubio and coworkers~\cite{PellegriniPhysRevLett2015, FlickACSPhoton2018} proposed an extension of optimized effective potentials (OEPs). 
However, the accuracy of this functional still needs to be assessed. 

Coupled cluster theory 
is one of the most successful methods for treating electron correlation in both ground and excited states of molecular systems~\cite{Bartlett2007,HelgakerBook2000}.
It is nowadays routinely applied to compelling chemical problems
due to major advances in computational resources and several decades of algorithmic developments.
Coupled cluster methods are available in several programs; we use the electronic structure program $e^\mathcal{T}$~\cite{eTpaper} for our developments.
In this paper, we extend
coupled cluster theory 
to treat strongly interacting electron-photon systems in a non-relativistic QED framework. 
We refer to the resulting method as quantum electrodynamics coupled cluster (QED-CC) theory.
To the best of our knowledge, 
this is the first coupled cluster formalism that incorporates many-body electron-photon operators for an \emph{ab initio} Hamiltonian.
Recently, a different coupled cluster formalism was  proposed by Mordovina \emph{et al}.~\cite{MordovinaCCArXiv2019}. Their study was limited to model Hamiltonians and they used state-transfer operators, instead of many-body operators, to describe the photonic part of the wave function. 
We should also mention that studies using other electronic structure methods and model or semi-empirical Hamiltonians have been presented by other authors~\cite{FregoniNatureCommun2018, FregoniChem2020, KowalewskiJChemPhys2016}.

In addition to presenting the complete formulation and implementation of QED-CC, we consider some interesting applications in photochemistry. In particular, 
we demonstrate how light-induced charge transfer in small dye molecules, commonly used as prototypes
for photovoltaic applications, can be modified by the quantized electromagnetic field. 
Furthermore, we show that
the presence of the cavity 
can break molecular symmetry and change relaxation mechanisms. Suitably defined fields can induce significant changes in both ground and excited state properties.
These results pave the way for novel strategies to  control photochemical reaction paths. 

\section{Coupled cluster theory for electrons}
\label{sec:CC-electrons}
In this section we introduce notation and important concepts needed to develop the electron-photon interaction model.
For a complete outline of coupled cluster theory, we refer to Ref.~\cite{HelgakerBook2000}.
In standard coupled cluster theory for singlet states, the many-body wave function is expressed using the exponential parametrization,
\begin{equation}\label{eq:CC_ansatz}
    \lvert \text{CC}\rangle = \exp(T) \lvert \text{HF}\rangle,
\end{equation}
where $\lvert \text{HF}\rangle$ is a reference wave function that is usually
chosen to be the 
closed-shell Hartree-Fock (HF) determinant. 
The cluster operator $T$ generates electronic excitations when operating on the reference and, in this way, the exponential produces a superposition of Slater determinants.
In the case of purely electronic many-body states, the cluster operator is defined as
\begin{equation}\label{eq:cluster_op}
    T = T_1 + T_2 + \dots + T_{N_e},
\end{equation}
where $N_e$ is the number of electrons. Each term corresponds to excitations (single, double, triple, and so on), i.e.
\begin{align}
    &T_1 = \sum_{ai} t_{ai} E_{ai} \label{eq:T1} \\
    &T_2 = \frac{1}{2}\sum_{aibj} t_{aibj} E_{ai}E_{bj}, \label{eq:T2}
\end{align}
where $E_{pq}=a^{\dagger}_{p\alpha}a_{q\alpha}+a^{\dagger}_{p\beta}a_{q\beta}$ (here, $a$ and $a^\dagger$ denote fermionic operators and ($\alpha$, $\beta$) denote spin projections) 
are singlet one-electron operators and the parameters $t_{ai}$ and $t_{aibj}$ are called cluster amplitudes. Furthermore, we let indices $(i,j,k,l)$ and $(a,b,c,d)$ label occupied and virtual HF orbitals, respectively. General orbitals are labeled $(p,q,r,s)$. The cluster operator can be expressed as
\begin{equation}
    T = \sum_\mu t_\mu \tau_\mu,
\end{equation}
where the excitation operators $\tau_\mu$ generate an orthonormal set of excited configurations:
\begin{equation}
    |\mu \rangle = \tau_\mu |\text{HF} \rangle.
\end{equation}
Together with $\vert \mathrm{HF} \rangle$, these configurations define a subspace of the Hilbert space in which the Schr{\"o}dinger equation is solved.
Inserting the coupled cluster wave function in Eq.~\eqref{eq:CC_ansatz} into the time-independent Schr{\"o}dinger equation, we obtain
\begin{equation}\label{eq:CC_equation}
    H_{e}\lvert \text{CC}\rangle = \lvert \text{CC}\rangle E_{\mathrm{CC}},
\end{equation}
where $H_{e}$ is the electronic Born-Oppenheimer Hamiltonian
\begin{equation}\label{eq:H_e}
    H_{e} = \sum_{pq}h_{pq}E_{pq} + \frac{1}{2}\sum_{pqrs}g_{pqrs}e_{pqrs} + h_{\text{nuc}}.
\end{equation}
The quantities $h_{pq}$ and $g_{pqrs}$ are one- and two-electron integrals respectively and, for convenience, we have introduced the operator
\begin{equation}\label{eq:identity}
e_{pqrs}=E_{pq}E_{rs}-\delta_{qr}E_{ps}=\sum_{\sigma,\tau=\alpha,\beta} a^{\dagger}_{p\sigma}a^{\dagger}_{r\tau}a_{s\tau}a_{q\sigma}.
\end{equation}
In coupled cluster theory,  Eq.~(\ref{eq:CC_equation}) is projected onto the set $\{ \vert \text{HF} \rangle, \vert \mu \rangle \}$. Consequently, the coupled cluster energy is given by
\begin{align}\label{eq:CCsolveq_1}
    E_{\mathrm{CC}} &= \langle\text{HF}\rvert \bar{H}_{e} \lvert \text{HF}\rangle
\end{align}
and the cluster amplitudes are determined from the equations
\begin{align}\label{eq:CCsolveq_2}
    \Omega_\mu &= \langle \mu \rvert \bar{H}_{e} \lvert \text{HF}\rangle = 0.
\end{align}
Here we have introduced the similarity transformed Hamiltonian
\begin{equation}\label{eq:H_bar}
     \bar{H}_e = \exp(-T) H_e \exp(T).
\end{equation}

Coupled cluster theory is equivalent to exact diagonalization of the Hamiltonian in Eq.~\eqref{eq:H_e}, also called full configuration interaction (FCI), when the cluster operator in Eq.~\eqref{eq:cluster_op} is untruncated and contains all possible excitations. When the excitation space is truncated, we obtain different levels of approximation 
and, needless to say, reduced 
computational cost. 
For example, 
$T=T_1$ defines the coupled cluster singles model (CCS) and 
$T=T_1+T_2$ 
the coupled cluster singles and doubles model (CCSD). Coupled cluster theory is manifestly size-extensive, also in its truncated forms, a property that ensures that the total energy of non-interacting subsystems is the sum of the subsystem energies. This is unlike similar truncation in configuration interaction theory, where extensivity errors can become arbitrary large for an increasing number of subsystems. 

Another important feature of coupled cluster theory is the size-intensivity of excitation energies: for non-interacting subsystems,
excitation energies
in
each subsystem do not change with the total system size~\cite{KochJChemPhys1990}.
The 
excitation energies are the eigenvalues of the Jacobian matrix 
\begin{equation}\label{eq:jacobian}
    A_{\mu \nu}= \frac{\partial \Omega_\mu}{\partial t_\nu} = \langle \mu \rvert [\bar{H}_{e}, \tau_\nu] \lvert \text{HF}\rangle.
\end{equation}
\noindent
Since this matrix is non-Hermitian, special attention is required at electronic degeneracies; at such points, the matrix can become defective/non-diagonalizable. Defects are therefore expected close to conical intersections, as
discussed in more detail in Section~\ref{sec:technical-aspects}. 

In coupled cluster theory there are two prevailing approaches to electronic excited states. One is coupled cluster response theory (CCRT), 
which
is based on a time-dependent formalism~\cite{KochJChemPhys1990}. 
This theory provides both size-extensive and size-intensive molecular properties, such as excitation energies and transition moments. The other is based on a time-independent formalism 
and is 
referred to as equation of motion coupled cluster (EOM-CC) theory~\cite{StantonJChemPhys1993}. In EOM-CC theory, the excitation energies are the same as in CCRT. However, some molecular properties are not guaranteed to scale correctly with system size. For instance, transition moments are not necessarily size-intensive~\cite{KochJChemPhys1994}. 
For the purpose of the present developments, which mainly relates to ground and excited state energies, the EOM-CC formalism is sufficient.
In EOM-CC, the similarity transformed Hamiltonian is expressed in the basis $\{ \vert \mathrm{HF} \rangle, \vert \mu \rangle \}$,
\begin{align}
\label{eq:eomH}
\begin{split}
    \bar{\bf{H}}&= 
    \begin{pmatrix}
     \langle \text{HF}| \bar{H}_e |\text{HF} \rangle & \langle \text{HF}| \bar{H}_e |\nu \rangle \\
     \langle \mu| \bar{H}_e |\text{HF} \rangle &  \langle \mu | \bar{H}_e |\nu \rangle
    \end{pmatrix}\\
    &=
    \begin{pmatrix}
     E_{\mathrm{CC}} & \eta_{\nu} \\
     0 &  A_{\mu \nu} + \delta_{\mu \nu}E_{\mathrm{CC}}
    \end{pmatrix}.
\end{split}
\end{align}
The left and right eigenvectors of $\bar{\mathbf{H}}$ define the excited state vectors, $\langle \mathcal{L}_k \vert$ and $\vert \mathcal{R}_k \rangle$, and the eigenvalues of $\bar{\mathbf{H}}$ are the energies of the states. We have used Eqs.~\eqref{eq:CCsolveq_1}, \eqref{eq:CCsolveq_2}, and \eqref{eq:jacobian} in the last equality. 
To extend coupled cluster theory to electron-photon systems, we need to introduce a new parametrization of the cluster operator. 

\section{Coupled cluster theory for electron-photon systems}

To describe the interaction of the electromagnetic field with atoms, molecules, and condensed matter systems, the low-energy limit of QED is usually sufficient~\cite{CohenTannoudjiBook1989, CraigBook1984}.
In particular, the non-relativistic Pauli-Fierz Hamiltonian in the dipole approximation~\cite{CohenTannoudjiBook1989,SpohnBook2004, RokajJPhysB2018, RuggenthalerNatureRevChem2018},
\begin{align}\label{eq:QEDHam}
\begin{split}
H_{\mathrm{PF}} &= H_e + \sum_{\alpha} \Bigl( \omega_{\alpha}b^\dagger_\alpha b_\alpha + \frac{1}{2} ({\boldsymbol{\lambda}}_{\alpha} \cdot \boldsymbol{d})^2
\\ 
&- \sqrt{\frac{\omega_{\alpha}}{2}}({\boldsymbol{\lambda}}_{\alpha} \cdot \boldsymbol{d})(b^\dagger_\alpha +b_\alpha) \Bigr),
\end{split}
\end{align}
is usually an accurate starting point for describing electronic systems in optical cavities.
The Hamiltonian in Eq.~(\ref{eq:QEDHam}) is expressed in Coulomb and length gauge \cite{CohenTannoudjiBook1989, CraigBook1984, SpohnBook2004}. We use the Born-Oppenheimer approximation and keep the nuclear positions fixed. 
The second term in the Hamiltonian is the purely photonic part, represented by a sum of harmonic oscillators, one for each frequency. We have neglected the zero-point energies. 
The operators $b^{\dagger}_{\alpha}$ and $b_{\alpha}$ are bosonic creation and annihilation operators, respectively. 
The third term is the dipole self-energy term, which ensures that the Hamiltonian is bounded from below~\cite{RokajJPhysB2018} and independent of origin. 
The last term, the bilinear coupling, 
couples the electronic and photonic degrees of freedom.
In the length gauge, the light-matter coupling is via the dipole operator 
\begin{equation}\label{eq:dipole}
\Vec{d} = \sum_{pq} \Vec{d}_{pq} E_{pq} \quad \Vec{d}_{pq} = \langle p| \vec{d}_{e} + \frac{\Vec{d}_{\text{nuc}}}{N_e} |q\rangle
\end{equation}
which consists of an electronic and a constant nuclear contribution, $\vec{d}_{e}$ and $\vec{d}_{\text{nuc}}$. The elements $\Vec{d}_{pq}$ denotes one-electron dipole integrals.
The coupling is described through the transversal polarization vector $\Vec{e}$ multiplied by the coupling strength $\lambda_{\alpha}$:
\begin{align}
    \Vec{\lambda}_{\alpha}=\lambda_{\alpha}\Vec{e}, \quad \lambda_{\alpha} =
\sqrt{\frac{1}{\varepsilon_0\varepsilon_r V_{\alpha}}}.
\end{align}
Here,
$\varepsilon_0$ and $\varepsilon_r$ are the permittivities of the vacuum and the dielectric materials separating the cavity mirrors, respectively~\cite{LatiniNanoLett2019}. The $\alpha$ mode quantization volume is denoted $V_\alpha$.

The dipole approximation is usually valid when the wavelength of the electromagnetic field is significantly larger than the size of the electronic system. There are, however, cases where the dipole approximation is not sufficiently accurate. For instance, this occurs when the size of the system is comparable to the cavity wavelength, or when the matter interacts with a
circularly or elliptically polarized field.
These aspects will be discussed in a forthcoming paper.

\subsection{\label{sec:QED-HF} The QED-HF method}

In order to formulate QED-CC, we need to define a suitable reference wave function.
We formulate an extension of the Hartree-Fock method to QED, hereafter referred to as QED-HF. 
The non-correlated electrons and photons in QED-HF are described by
\begin{align}\label{eq:QEDHF_ansatz}
    \lvert R\rangle &=
    \lvert \text{HF}\rangle \otimes \lvert P \rangle
    \\
    \lvert P \rangle &=
    \sum_{\vec{n}}  
    \prod_{\alpha} (b^{\dagger}_\alpha)^{n_\alpha}
    \lvert 0 \rangle c_{\vec{n}}
\end{align}
where $\lvert 0\rangle$ is the photon vacuum 
and $c_{\vec{n}}$ are expansion coefficients for the photon number states. 
The coefficient $c_{\Vec{n}}$, where $\vec{n} = (n_1,n_2,\dots)$, corresponds to the state with $n_\alpha$ photons in mode $\alpha$.
Starting from Eq.~\eqref{eq:QEDHF_ansatz}, and assuming that $\vert R \rangle$ is normalized,
the energy,
\begin{align}
   E_\mathrm{QED-HF} = \langle R\rvert H\lvert R\rangle,
\end{align}
is minimized with respect to Hartree-Fock orbitals and photon coefficients $c_{\vec{n}}$. Note that 
QED-HF
can be considered
a special case of QEDFT 
with an appropriately chosen exchange-correlation functional.

For a given Hartree-Fock state, the energy can be minimized with respect to the photon coefficients. This can be achieved by diagonalizing the photonic Hamiltonian
\begin{align}\label{eq:phot_ham}
\begin{split}
    \langle H_\mathrm{PF} \rangle &= E_\mathrm{HF} + \sum_{\alpha} \Bigl(
    \omega_{\alpha}b^{\dagger}_{\alpha}b_{\alpha} +\frac{1}{2} \langle (\Vec{\lambda}_\alpha \cdot \Vec{d} )^2 \rangle \\
    &-\sqrt{\frac{\omega_{\alpha}}{2}} (\Vec{\lambda}_\alpha \cdot \langle \Vec{d} \rangle) (b^{\dagger}_{\alpha}+b_{\alpha})
    \Bigr), 
\end{split}
\end{align}
where the mean value is with respect to $\vert \mathrm{HF} \rangle$. This Hamiltonian can be diagonalized by the unitary coherent state transformation~\cite{klauder1985coherent} 
\begin{align}\label{eq:b-unitary}
    U(\Vec{z}) = \prod_\alpha \exp(z_{\alpha}b_{\alpha}^{\dagger} - z_{\alpha}^* b_{\alpha})
\end{align}
with a suitable choice of $\Vec{z}$. In the transformed basis, the photonic Hamiltonian is
\begin{align}\label{eq:sim_trans_H}
\begin{split}
    \langle H_{\mathrm{PF}} \rangle_{\vec{z}} &= E_{\mathrm{HF}} + \sum_{\alpha} \Bigl(
    \omega_{\alpha} (b^{\dagger}_{\alpha} + z^*_{\alpha}) (b_{\alpha} + z_{\alpha}) + \frac{1}{2} \langle(\Vec{\lambda}_{\alpha} \cdot \Vec{d} )^2\rangle \\
    &- \sqrt{\frac{\omega_{\alpha}}{2}} (\Vec{\lambda}_\alpha \cdot \langle \Vec{d} \rangle) (b^{\dagger}_{\alpha} + b_{\alpha} + z_{\alpha} + z_{\alpha}^*)
    \Bigr).
\end{split}
\end{align}
If we choose
\begin{align}
    (\vec{z}_0)_{\alpha} =-\frac{\Vec{\lambda}_\alpha \cdot \langle\Vec{d}\rangle}{\sqrt{2\omega_{\alpha}}}, \label{eq:z_coherent}
\end{align}
the Hamiltonian reduces to
\begin{align}
    \langle H_{\mathrm{PF}} \rangle_{\vec{z}_0} &= E_{\mathrm{HF}} + \frac{1}{2}\sum_\alpha \langle (\Vec{\lambda}_\alpha \cdot (\Vec{d} - \langle\Vec{d}\rangle))^2\rangle + \sum_{\alpha} 
    \omega_{\alpha} b^{\dagger}_{\alpha}b_{\alpha}.
\end{align}
The eigenvectors of this operator are the photon number states, and the lowest eigenvalue corresponds to the vacuum state. In the untransformed basis, that is, for the Hamiltonian in Eq.~\eqref{eq:phot_ham}, the eigenstates are the generalized coherent states
\begin{align}
    |z_{\alpha},n_{\alpha}\rangle = \exp(z_{\alpha}b_{\alpha}^{\dagger} - z_{\alpha}^* b_{\alpha}) |n_{\alpha}\rangle
\end{align}
where $z_\alpha$ is given in Eq.~\eqref{eq:z_coherent} and
\begin{equation}
    |n_{\alpha}\rangle = \frac{(b^\dagger_{\alpha})^{n_{\alpha}}}{\sqrt{n_{\alpha}!}} |0\rangle
\end{equation}
are the normalized photon number states for mode $\alpha$.

Applying the unitary transformation to the original Pauli-Fierz Hamiltonian, Eq.~\eqref{eq:QEDHam}, 
gives us the final expression for the Hamiltonian in the coherent state basis: 
\begin{align}\label{eq:sim_trans_H_final}
\begin{split}
    H &= H_{e} + \sum_{\alpha}\Bigl(\omega_{\alpha} b^{\dagger}_{\alpha}b_{\alpha}+\frac{1}{2} \bigl(\Vec{\lambda}_{\alpha} \cdot(\Vec{d}-\langle\Vec{d}\rangle)\bigr)^2 \\
    &- \sqrt{\frac{\omega_{\alpha}}{2}}\bigl( \vec{\lambda}_{\alpha} \cdot (\Vec{d}-\langle\Vec{d}\rangle) \bigr) (b^{\dagger}_{\alpha} + b_{\alpha})\Bigr).
\end{split}
\end{align}
This Hamiltonian will be used in QED-CC.
Note that the operator is manifestly origin invariant, differently from Eq.~\eqref{eq:QEDHam}, where the invariance is obtained through a gauge transformation.
This is also true for charged systems, where the dipole moment operator depends on the choice of origin.
In the coherent state basis, the bilinear coupling and self-energy terms depend on fluctuations of the dipole moment away from the mean value. 
We point out that the nuclear dipole gives no contributions to the Hamiltonian. 
However, note that the dependence on the nuclear dipole now resides in the wave function through the coherent state transformation, see Eq.~\eqref{eq:z_coherent}.

As we have shown, the QED-HF reference state is now given by
\begin{equation}\label{eq:qed-hf-gs}
    \lvert R \rangle = \lvert \text{HF} \rangle \otimes \lvert 0 \rangle.
\end{equation}
The Hartree-Fock equations are solved using standard techniques~\cite{pulay1982, HelgakerBook2000}. In every iteration, the Hartree-Fock orbitals are updated and used to evaluate $\Vec{z}_0$, see Eq.~\eqref{eq:z_coherent}. The inactive Fock matrix~\cite{HelgakerBook2000} used in the optimization is given by
\begin{eqnarray}\label{eq:fock-matrix}
    F_{pq} &=& F_{pq}^e + \frac{1}{2} \sum_{\alpha}  \left(
    \sum_a (\Vec{\lambda}_\alpha \cdot \Vec{d}_{pa}) (\Vec{\lambda}_\alpha \cdot \Vec{d}_{aq}) \right. \nonumber\\ 
    &-& \left.\sum_i (\Vec{\lambda}_\alpha \cdot \Vec{d}_{pi}) (\Vec{\lambda}_\alpha \cdot \Vec{d}_{iq})
    \right),
\end{eqnarray}
where ${\bf F}^{e}$ is the 
standard closed-shell
electronic Fock matrix~\cite{HelgakerBook2000} and the stationary condition is equivalent to $F_{ia} = 0$. 
The QED-HF ground state energy can now be written as
\begin{align}
\begin{split}\label{eq:qed-hf-energy}
    E_{\text{QED-HF}} &= E_{\text{HF}} + \frac{1}{2} \sum_\alpha  \left\langle (\Vec{\lambda}_\alpha \cdot (\Vec{d}-\langle\Vec{d}\rangle))^2 \right\rangle\\
    &= E_{\text{HF}} + \sum_{\alpha,ai} (\Vec{\lambda}_{\alpha} \cdot \Vec{d}_{ai})^2,
\end{split}
\end{align}
where the correction to the electronic energy 
can be understood as the variance in the dipole interacting with the photon field. 
For an infinite number of modes, we expect, as is standard in QED \cite{weinberg1995quantum}, to encounter divergencies due to the second term of Eq.~\eqref{eq:qed-hf-energy}. 

We should point out that the eigenvalues of $\bf{F}$, normally interpreted as orbital energies, are origin dependent for charged systems. As a consequence, 
applying concepts that depend on the orbital energies, like the Koopmans' theorem~\cite{koopmans1933ordering}, will require a different choice of the occupied-occupied and virtual-virtual blocks of $\bf{F}$.
For the same reason, $\bf{F}$ cannot be used as a zeroth order Hamiltonian in perturbation theories such as CC2~\cite{Christiansen1995CC2} and CC3~\cite{Christiansen1995,Koch1997}. 
An origin independent $\bf{F}$ can be obtained 
by performing an appropriate unitary transformation \cite{Dalgaard_1976} that mixes the electronic and photonic degrees of freedom. In this way, the dressed electrons have well-defined orbital energies also for charged systems. This will be considered in a future publication. Note, however, that the origin dependence of the eigenvalues of $\bf{F}$ does not imply a loss of origin invariance in QED-HF (see Appendix~\ref{app:QED-HF}).

\subsection{The QED-CC method}\label{sec:QED-CC}

Extending the exponential parametrization in Eq.~\eqref{eq:CC_ansatz} to QED requires that the cluster operator generates excitations both in the purely photonic and electron-photon coupling spaces~\cite{MordovinaCCArXiv2019}. The cluster operator can therefore be partitioned as
\begin{equation}\label{eq:QED_cluster_operator}
    T = T_{e} + T_{p} + T_{int},
\end{equation}
where $T_{e}$ is the standard cluster operator for the electrons and $T_{p}$ and $T_{int}$ consist of photon and electron-photon operators, respectively.

The purely photonic operator is defined as
\begin{eqnarray}\label{eq:phot_cluster}
    T_{p} &=& \sum_{\vec{n}} \Gamma^{\vec{n}} = \sum_{\vec{n}} \gamma^{\vec{n}} \prod_\alpha (b^{\dagger}_{\alpha})^{n_\alpha} .
\end{eqnarray}
In this equation, $\gamma^{\vec{n}}$ are photon amplitudes. 
The form of the photonic operator was chosen to expand the photonic part of the Hilbert space and to give commuting cluster operators (as in the electronic case). Since $\exp(T_{p})$ does not terminate, the parametrization is able to incorporate many-body effects within the limitation imposed by the projection space. In contrast, Mordovina \emph{et al}.~\cite{MordovinaCCArXiv2019} used a nilpotent photonic operator that only enters linearly in the expansion of the coupled cluster state.

The excitations in the electron-photon interaction operator $T_{int}$ are defined as direct products of electronic and photonic excitations. Thus, the operator can be expressed as
\begin{equation}\label{eq:Coup_cluster_op}
    T_{int} = \sum_{\vec{n}} S^{\vec{n}}_1 + S^{\vec{n}}_2 + \dots + S^{\vec{n}}_{N_e}
\end{equation}
where, for instance,
\begin{align}
    S^{\vec{n}}_1 &= \sum_{ai} s^{\vec{n}}_{ai} E_{ai} \prod_\alpha (b^\dagger_\alpha)^{n_\alpha}
    \\
    S^{\vec{n}}_2 &= \frac{1}{2} \sum_{aibj} s^{\vec{n}}_{aibj} E_{ai}E_{bj} \prod_\alpha (b^\dagger_\alpha)^{n_\alpha}.
\end{align}
The new cluster amplitudes, $\gamma^{\Vec{n}}, s^{\vec{n}}_{ai}$, $s^{\vec{n}}_{aibj}$, etc., are 
parameters that will be determined from a set of projection equations. 

Hierarchies of approximations are formulated by truncating the cluster operator and the associated projection space. Here we implement CCSD with the special case of a single photon mode, where we only include one photon in the cluster operator. The coupled cluster wave function is then given by
\begin{align}
    \vert \mathrm{CC} \rangle = \exp(T) \vert \mathrm{R} \rangle,
\end{align}
where $\vert \mathrm{R} \rangle$ is the QED-HF reference given in Eq.~\eqref{eq:qed-hf-gs}, and the cluster operator is
\begin{align}
    T = T_1 + T_2 + \Gamma^1 + S_1^1 + S_2^1. \label{eq:QED_CCSD_T}
\end{align}
Electronic excitations are described at the singles and doubles level, and the photon mode is coupled to these excitations through $S_1^1$ and $S_2^1$, respectively. The electronic operators $T_1$ and $T_2$ are given in Eqs.~\eqref{eq:T1} and \eqref{eq:T2}. The photon and electron-photon operators are defined as
\begin{align}
    \Gamma^1 &= \gamma b^\dagger \\
    S_1^1 &= \sum_{ai} s_{ai} E_{ai} b^\dagger \\
    S_2^1 &= \frac{1}{2}\sum_{aibj} s_{aibj} E_{ai} E_{bj} b^\dagger.
\end{align}
This model is referred to as QED-CCSD-1 with one photon mode, where ``1'' refers to the photonic excitation order. More involved terminology will be required to describe the full hierarchy. In the notation used by Mordovina \emph{et al}.~\cite{MordovinaCCArXiv2019}, this is a QED-CC-SD-S-DT model. However, due to the difference in photonic excitation operators 
and the coherent state basis, 
the model described here is not directly comparable to the one in Ref.~\cite{MordovinaCCArXiv2019}.
Even if only one photon creation operator is included, the exponential will partially incorporate two photon contributions into the wave function. Thus we expect the convergence with respect to photons will be faster than CI-like diagonalization in photon number states. Furthermore, notice that the generalized coherent states basis incorporates higher photonic excitations as well.

The projection space used in Eq.~\eqref{eq:CCsolveq_2} is defined from the excitations included in the cluster operator. With the notation
\begin{align}
    \vert \mathrm{HF}, n \rangle &= \vert \mathrm{HF} \rangle \otimes \vert n \rangle \\
    \vert \mu, n \rangle &= \vert \mu \rangle \otimes \vert n \rangle,
\end{align}
the projection basis is
\begin{align}
    \{ \vert \mathrm{HF}, 0 \rangle, |\mu, 0\rangle, |\mu,1\rangle, \lvert \text{HF},1\rangle \}, \label{eq:QED_CCSD_projection}
\end{align}
where $\vert \mathrm{HF}, 0 \rangle = \vert R \rangle$ and $\mu$ is restricted to single and double excitations. The derivation of the amplitude equations follows the same procedure as in the electronic case, and the truncation of the equations is determined by the projection space and 
the commutator expansion of the similarity transformed Hamiltonian.
Explicit formulas
are presented in Appendix~\ref{app:QED-CC-SD-S-DT}.

The formation of polaritons usually appears in the optical spectrum as a Rabi splitting (proportional to the coupling strength $\lambda$) of the electronic states due to the coupling to the quantum field. Hence, we must also describe the excited states of the coupled system.
In coupled cluster theory, electronic excitation energies may be determined using EOM-CC theory, as described in Section~\ref{sec:CC-electrons}. 
The projection space in QED-CCSD is extended, relative to the electronic case, giving rise to additional blocks in the Jacobian matrix: 
\begin{align}\label{eq:qed-jacobian-matrix}
    \bf{A} = \begin{pmatrix} {\bf{A}}_{e,e} &  {\bf{A}}_{e,ep} & {\bf{A}}_{e,p} \\
    {\bf{A}}_{ep,e} & {\bf{A}}_{ep,ep} & {\bf{A}}_{ep,p} \\
    {\bf{A}}_{p,e} & {\bf{A}}_{p,ep} & {\bf{A}}_{p,p}
    \end{pmatrix}.
\end{align}
In addition to the electronic Jacobian ${\bf{A}}_{e,e}$, there are coupling blocks between electronic ($e$), electronic-photonic ($ep$), and photonic ($p$) configurations; see Eqs.~\eqref{eq:QED_CCSD_T} and \eqref{eq:QED_CCSD_projection}. 
Explicit formulas for  the sub-blocks
of the Jacobian,
\begin{align}
    \label{eq:qed-cc-jaco}
    A_{\mu n,\nu m} = \langle\mu, n|[\bar{H}, \tau_{\nu}(b^{\dagger})^{m}] \lvert R \rangle,
\end{align}
are presented in Appendix~\ref{app:EOM-QED-CC-SD-S-DT}, along with the corresponding sub-blocks of the $\vec{\eta}$ vector.
The ground and excited state QED-CCSD equations are solved using standard methods~\cite{SCUSERIA1986236, HelgakerBook2000,DAVIDSON197587}.

To ensure a balanced description of the electron-photon states, it is important to use all product operators between the included electronic and photonic excitations in the cluster operator, $\tau_\mu (b^\dagger)^n$. For instance, suppose that the coupling is zero and consider the two states $|e,1\rangle$ and $|e,0\rangle$. 
As $\omega_{cav}$ tends to zero, these states will only become degenerate if $\Gamma^1$ is coupled to both $T_1$ and $T_2$. Neglecting $S_2^1$ would lead to unphysical results.

Additional properties of the 
electron-photon system can be calculated from the left and right eigenvectors  of $\bar{\bf{H}}$. For instance, we can evaluate
the ground and excited state electronic EOM density matrices as 
\begin{equation}\label{eq:cc_eom_density}
    D^k_{pq} = \langle \mathcal{L}_k \rvert \exp(-T) E_{pq} \exp(T) \lvert \mathcal{R}_k \rangle.
\end{equation}
For a description of other molecular properties, we refer the reader to the literature~\cite{StantonJChemPhys1993,KochJChemPhys1990}.

\begin{figure*}[ht!]
    \centering
    \includegraphics[trim={0.7cm 5.0cm 0.1cm 6cm},clip,width=\textwidth]{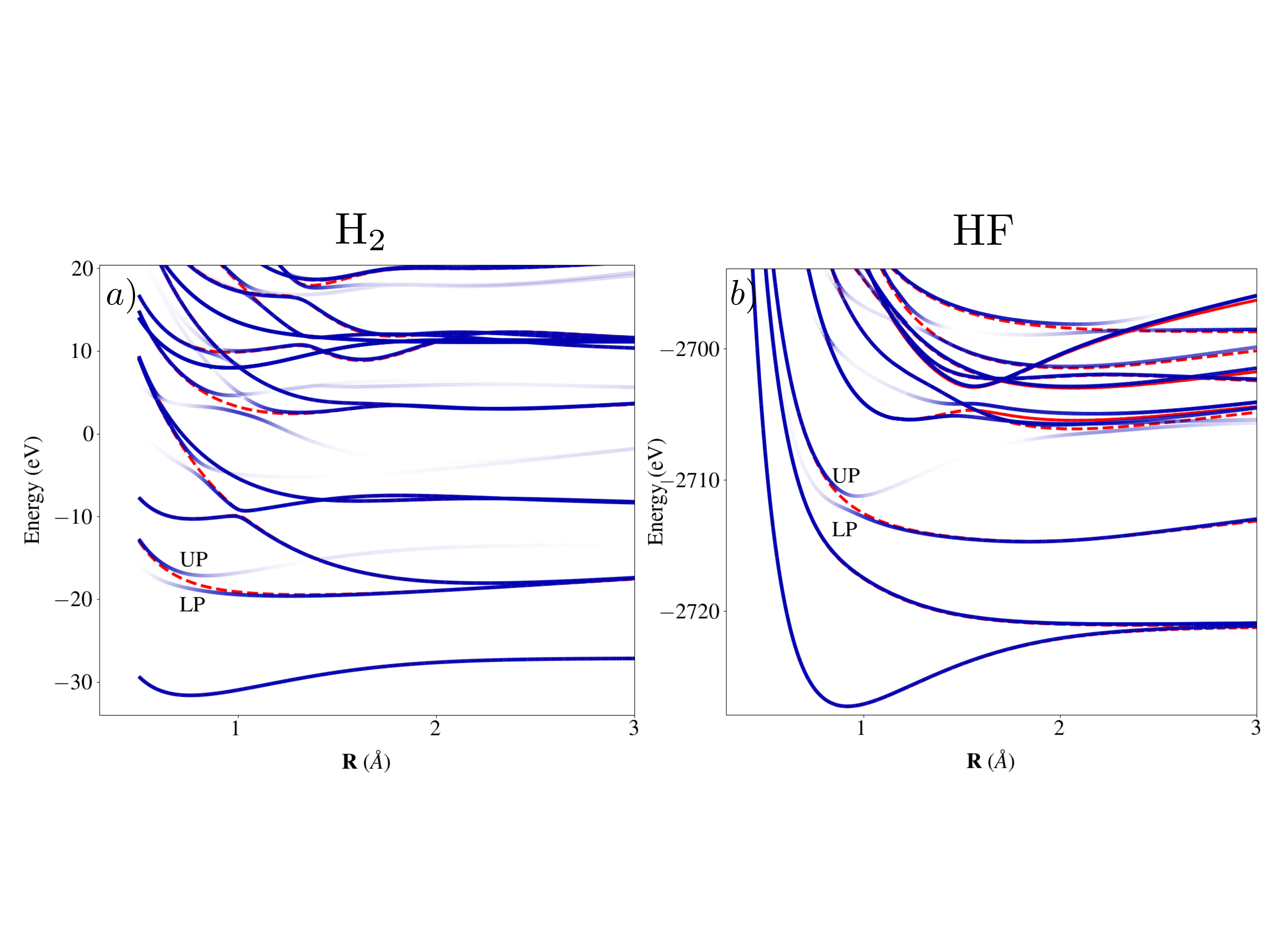}
    \caption{Potential energy curves calculated with (blue solid lines) and without (red dashed lines) an optical cavity for a) H$_2$ and b) HF. 
    The polarization is along the main axis of the molecules and the field is in resonance with the first bright excitation of the system at its equilibrium geometry. The coupling strength is set to $\lambda=0.05$ in both cases. The blue colormap indicates the electronic/photonic character of the states.}
    \label{fig:H2andHF}
\end{figure*}

\subsection{Some technical aspects}\label{sec:technical-aspects}

The Pauli-Fierz Hamiltonian in Eq.~\eqref{eq:QEDHam} is defined on the direct product Hilbert space $\mathscr{H} = \mathscr{H}_e \otimes \mathscr{H}_{p}$. In the truncated description, where $\mathscr{H}_e$ and  $\mathscr{H}_{p}$ are finite-dimensional, $\exp(T)\vert \mathrm{R} \rangle$ has an effectively finite expansion due to the finite projection basis. For instance, in QED-CCSD-1, the terms in $\exp(T)\vert \mathrm{R} \rangle$ that give non-zero contributions are up to quadruple electronic excitations and double photonic excitations. 
However, in the limit of an infinite-dimensional $\mathscr{H}_{p}$, 
special care might be required to define the exponential operator~\cite{Fisher1984}. 

Due to the non-Hermiticity of coupled cluster theory, it is known to give non-physical complex energies close to conical intersections between excited states of the same symmetry~\citep{hattig2005structure, Kohn2007, Kjonstad2017}. The same issues can arise in QED-CC and were mentioned by Mordovina \emph{et al}.~\citep{MordovinaCCArXiv2019}. 
The problem can be traced to 
defects in the Jacobian matrix
for a truncated
projection basis. 
In the untruncated case, the states satisfy generalized orthogonality relations, ensuring a correct description of conical intersections~\cite{Kjonstad2017}. To obtain a physically correct description with a truncated excitation space, one can enforce the orthogonality conditions 
\begin{align}
    \langle \mathcal{R}_k \vert \exp(T^\dagger) \mathcal{P}  \exp(T)  \vert\mathcal{R}_l \rangle = \delta_{kl},
\end{align}
where $\mathcal{P}$ is some projection operator~\citep{Kjonstad2017b, Kjonstad2019}. This approach, unlike \emph{a posteriori} corrections~\citep{Kohn2007,MordovinaCCArXiv2019}, can be extended to analytical energy gradients and nonadiabatic coupling elements using well-established Lagrangian techniques~\citep{helgaker1989configuration, Koch1990energyderiv}.
In passing, we note that no defects or complex eigenvalues were encountered in the results reported in this work.

\section{\label{sec:results} Molecular Polaritons}

In this section, the QED-CCSD-1 model is used to investigate cavity-induced effects on the chemistry of molecules. 
All calculations are performed using a development version of \textit{e$^T$}~\cite{eTpaper}.
Molecular geometries are provided in Supplemental Material~\cite{Supp_Mat}.
A single cavity mode is used throughout; this approximation typically breaks down for small values of $\omega_{cav}$,
when several replica states overlap energetically with electronic states. This 
can occur in large cavities, where $\omega_{cav}$ may be smaller than the electronic spectral range. In the following calculations, we assume that the single mode approximation is valid for the investigated properties.

In all calculations, we have chosen a coupling strength $\lambda = 0.05$. The coupling strength is usually calibrated using the Jaynes-Cummings model, where $\lambda = 0.05$ corresponds to a value well within the strong-coupling regime. Although this leads to relatively large Rabi splittings ($\sim$ 1 eV),comparable but smaller Rabi splittings have also been observed experimentally ($\sim$ 0.3 eV) \cite{ChikkaraddyNature2016, zhong2016non}.

\subsection{\label{sec:H2_HF} Diatomic molecules}

Interesting QED effects can be observed for small diatomic molecules, such as H$_2$ and HF. We also use these molecules to benchmark the coupled cluster model against the more accurate QED-FCI approach. The comparison shows an excellent agreement, see Appendix~\ref{app:comp_CC_FCI} for a detailed discussion.
For the calculations here we use a Gaussian basis set, in particular, cc-pVDZ~\cite{DunningJChemPhys1989}.

The potential energy curves for the ground and excited states of these systems are shown in Fig.~\ref{fig:H2andHF}.
The conical intersections and avoided crossings in the UV range define the photochemical properties of these molecules.
An optical cavity set in resonance with one of the excited states can completely restructure the excitation landscape and redefine the photochemistry of the system.
The color map in Fig.~\ref{fig:H2andHF} indicates the electronic/photonic contributions to the states.
The electronic states are highlighted in blue, while the photonic states are transparent white.
For more details, see Appendix~\ref{app:fractions}.

Considering first H$_2$ set in resonance with the first singlet excited state (at the ground state equilibrium geometry), we are able to induce significant changes of the potential energy curves.
Here we focus on the first Rabi splitting, where, in particular, the upper (UP) polariton is more bound than the bare electronic state.
Hence, it should be possible to trap the molecule in the UP state. In contrast, the bare electronic state has, to a larger extent, a dissociative character.

Similar conclusions can be drawn for the HF molecule. Differently from H$_2$, it has a permanent dipole moment. However, this seems to have minor effects on the general landscape. This is consistent with the fact that the total energy depends only on the fluctuations in the dipole moment (see Section~\ref{sec:QED-HF}).      

From the above analysis, we see that the application of a quantum field inside an optical cavity can be used to fine-tune the excited state properties of molecules. This opens the way towards new decay paths, the possibility of trapping systems in excited dark polaritons, and many other effects re-designing the molecular photochemistry.

\begin{figure}[ht!]
   \centering
   \includegraphics[trim={0cm 0cm 0cm 0cm},clip,width=8.6cm]{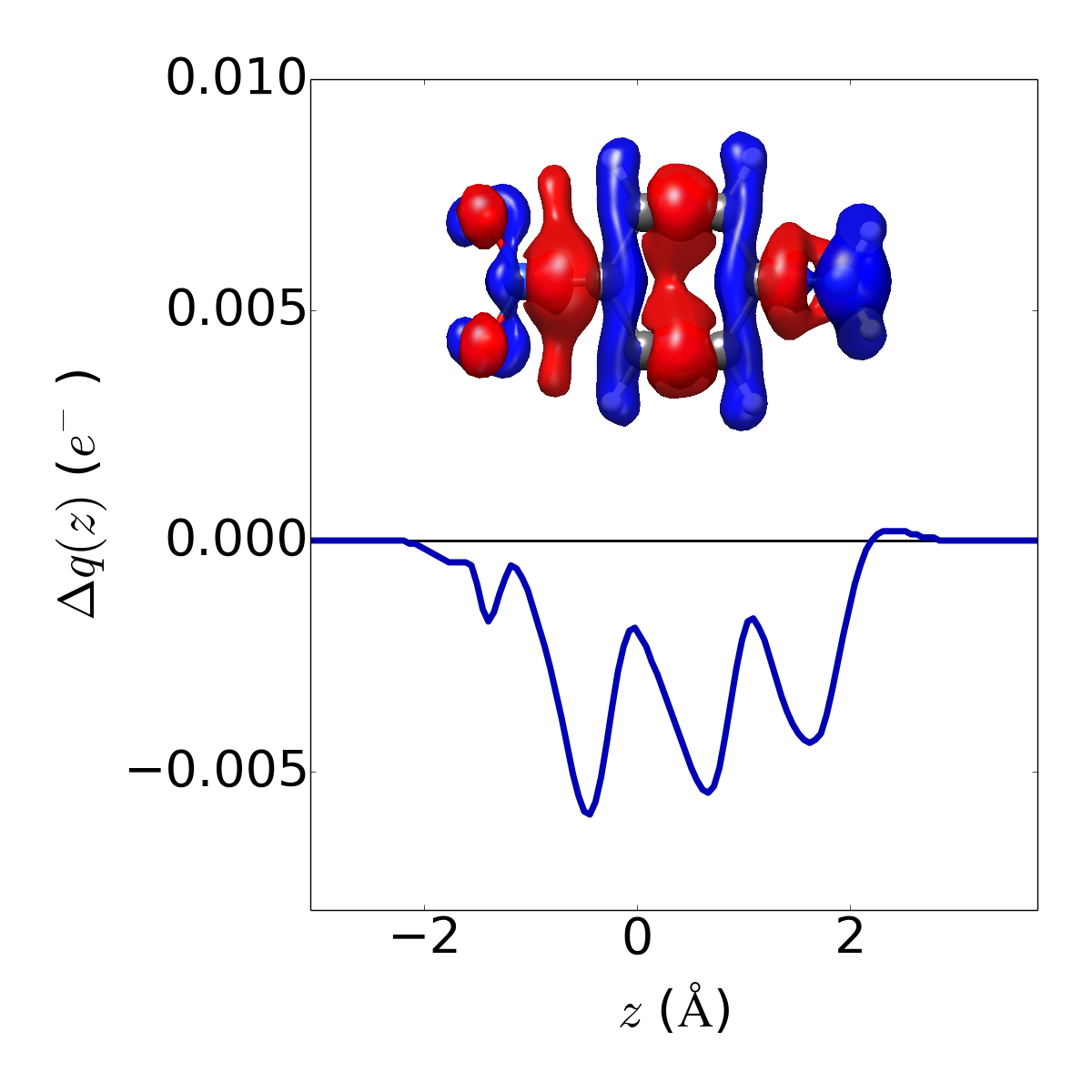}
   \caption{The QED-CCSD-1 ground state density difference induced in PNA by an optical cavity ($\lambda=0.05$ and $\omega_{cav}=$$4.84$~eV) and the corresponding charge displacement analysis. The blue/red regions represents charge accumulations/depletions. The isosurface value is $\pm 5\cdot 10^{-5}$ $e^-$a.u.$^3$}
   \label{fig:PNA_gs_dens}
\end{figure}
\subsection{Charge transfer molecules}

Recently, some research groups have suggested that quantum fields can have a significant impact on the charge transfer~\cite{SchaeferPNAS2019,MandalChemRxiv2020} and energy transfer~\cite{DuChemSci2018,SchaeferPNAS2019} properties in molecules. 
These preliminary studies are based on model Hamiltonians; thus, only qualitative interpretations of the phenomena are provided.
Here we present a quantitative analysis of cavity-induced effects on a charge transfer process.

We investigate p-nitroaniline (PNA), a simple amine often used as a prototype dye for solar energy applications, for instance in dye-sensitized solar cells~\cite{OReganNature1991, NazzeeruddinJAmChemSoc2005}.
This molecule has an intense low-lying charge transfer excitation (at about 3-4 eV) that potentially can be used to inject charge in a semiconductor and produce a current. Developing strategies to control the charge transfer process is of fundamental importance to increase photovoltaic efficiencies.

In our calculations on PNA, we have used the cc-pVDZ basis and oriented the polarization ($\lambda=0.05$) along the principal axis of the molecule. The molecular structure was optimized with DFT/B3LYP using the 6-31+G* basis set~\cite{HariharanTheorChimActa1973}.

\begin{figure}[ht!]
   \centering
   \includegraphics[trim={1.5cm 1.5cm 1.5cm 1.5cm},clip,width=8.6cm]{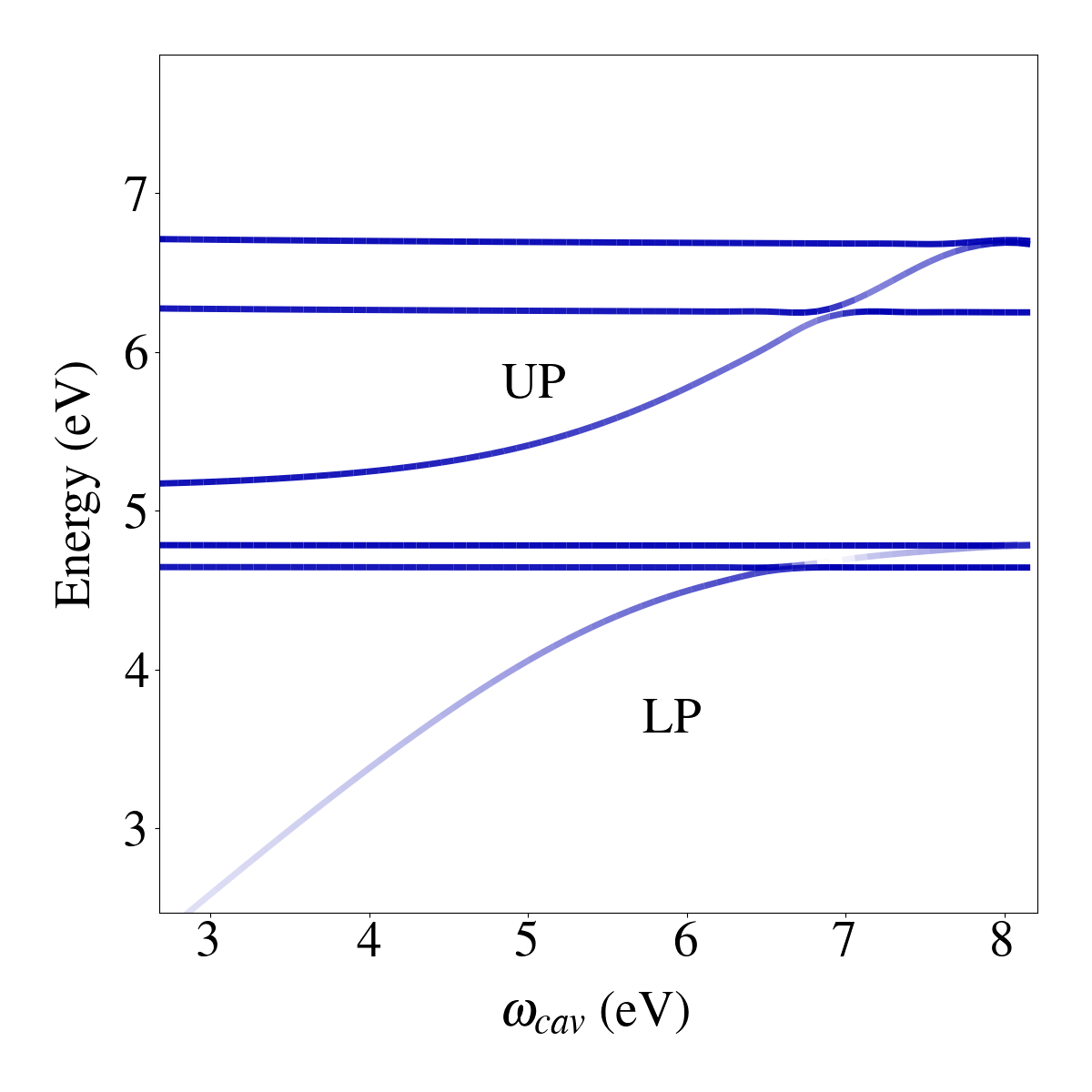}
   \caption{The QED-CCSD-1 dispersion with respect to cavity frequency $\omega_{cav}$ of the excitation energies in PNA. The blue colormap indicates the electronic/photonic character of the states.}
   \label{fig:PNA_omega_disp}
\end{figure}
\begin{figure}[ht!]
    \centering
    \includegraphics[trim={0cm 0cm 0cm 0cm},clip,width=8.6cm]{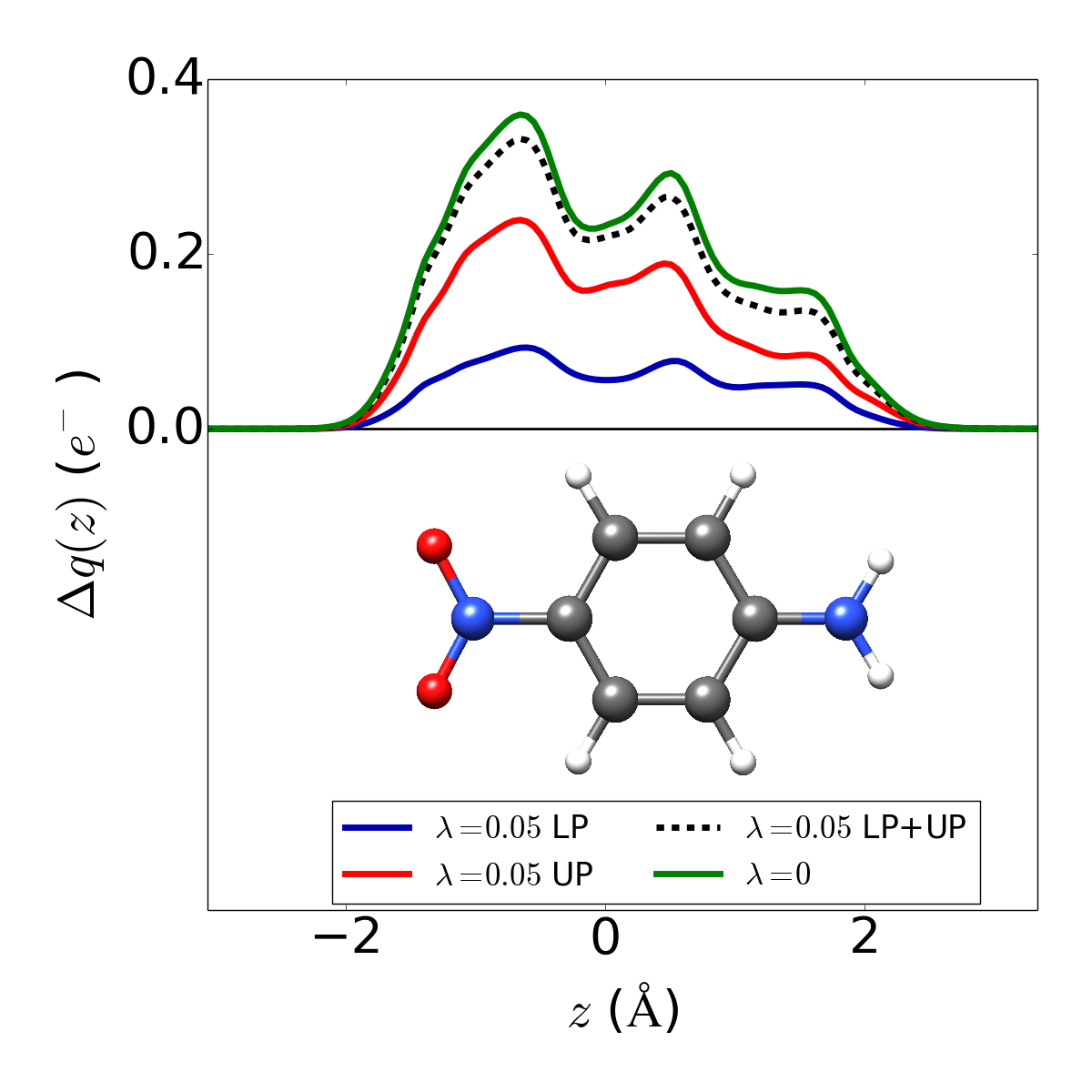}
    \caption{The QED-CCSD-1 excited state charge displacement analysis of PNA with and without an optical cavity ($\lambda=0.05$ and $\omega_{cav}=$ $4.84$~eV). The charge displacement functions with cavity for lower (LP - blue) and upper (UP - red) polaritons are shown. In green we show the function for the charge transfer state without the cavity. The dashed black line represents the sum of the curves for LP and UP.}
    \label{fig:PNA_ex_dens}
\end{figure}
\begin{figure*}[ht!]
    \centering
    \includegraphics[trim={0cm 0cm 0cm 0cm},clip,width=0.7\textwidth]{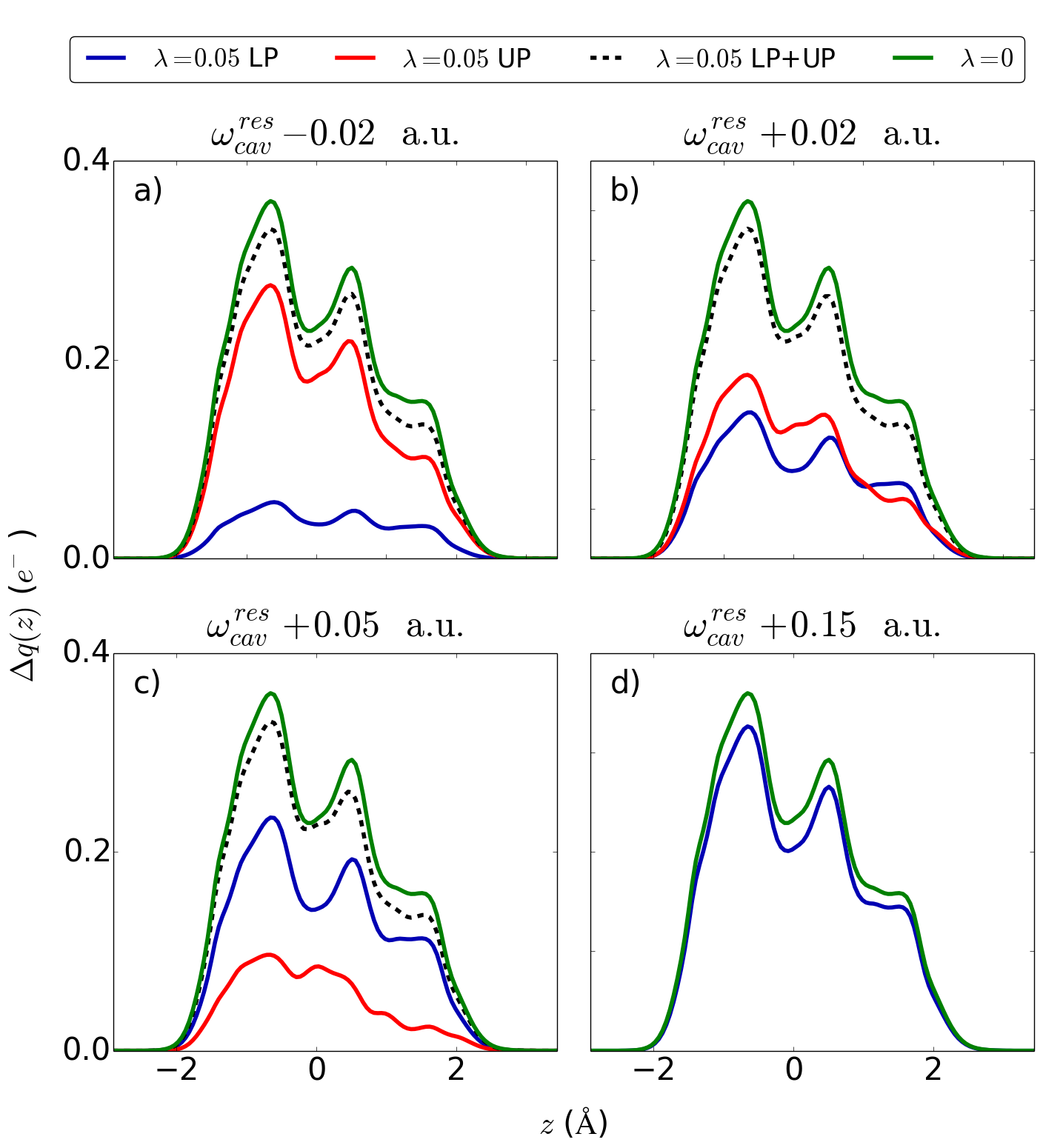}
    \caption{The QED-CCSD-1 excited state charge displacement analysis of PNA with and without an optical cavity ($\lambda=0.05$ and $\omega^{res}_{cav}=4.84$~eV). The charge displacement functions with cavity for lower (LP - blue) and upper (UP - red) polaritons are shown. In green we show the function for the charge transfer state without the cavity. The dashed black line represents the sum of the curves for LP and UP.}
    \label{fig:PNA_ex_dens_freq}
\end{figure*}
Initially we investigate the effects of the cavity on the electronic ground state by analyzing the electron density. This can conveniently be carried out using the charge displacement analysis~\cite{RoncaJChemPhys2014, BelpassiJAmChemSoc2008, CappellettiAccChemRes2012}. The charge displacement function is defined as
\begin{equation}\label{eq:CD}
\Delta q(z)=\displaystyle  \int_{-\infty}^{+\infty}
\int_{-\infty}^{+\infty} \int_{-\infty}^{z}  \Delta\rho(x,y,z') \, \mathrm{d}x \, \mathrm{d}y  \, \mathrm{d}z',
\end{equation}
where we integrate over the electron density difference $\Delta \rho$.
This function measures the amount of charge that has been moved along the $z$ coordinate.
In particular, if $\Delta q(z)$ is positive, charge is transferred from right to left, and if negative, charge is transferred in the opposite direction. 

In Fig.~\ref{fig:PNA_gs_dens}, we show the QED-CCSD-1 charge displacement function and isosurface for the ground state electron density difference with and without cavity, $\Delta\rho=\rho^{cav}_{gs}-\rho^{nocav}_{gs}$.
The cavity is set in resonance with the most intense low-lying charge transfer excited state, $\omega_{cav} = 4.84$ eV.
Although the charge displacement is small, a clear cavity-induced charge reorganization is observed in the ground state. Specifically, we have a charge transfer of about 0.005 $e^-$ going from the acceptor (NO$_2$) to the nitrogen atom of the donor (NH$_2$) group, reducing the magnitude of the dipole moment from $6.87$~D in a vacuum to $6.77$~D in the cavity.
This counter-intuitive effect is in agreement with previous QEDFT/OEP studies from Flick \textit{et al.}~\cite{FlickACSPhoton2018}. The cavity field accumulates more charge in the high-density regions. In this way, the variance of the dipole operator is reduced, see Eq.~\eqref{eq:qed-hf-energy}.

In the excited states of PNA, the cavity-induced effects are more evident. 
In Fig.~\ref{fig:PNA_omega_disp} we show the dispersion of the low-lying excitation energies of PNA with respect to the cavity frequency $\omega_{cav}$.
Due to the large transition dipole moment for the charge transfer excitation, a large Rabi splitting is observed when the cavity is resonant with this state. 
As discussed in the previous section, the cavity can induce significant changes in the excited states.
Specifically, we have a state inversion between the lower polariton and the first two excited states. This effect in PNA and other dye molecules could be important for photovoltaic applications, where a proper alignment of charge transfer states with the states in a semiconductor is essential to optimize solar cell efficiency. 

In Fig.~\ref{fig:PNA_ex_dens}, we show a charge displacement analysis of the charge transfer state, with a resonant quantum field.
Here we use the density difference between the ground and excited state, $\Delta\rho=\rho_{es}-\rho_{gs}$, with and without the cavity. Both the lower and upper polariton have charge transfer character and are shown separately.

Differently from the ground state, now a sizable charge transfer of almost $0.4$ $e^-$ are moved from the donor to the acceptor group.
The cavity divides the total charge transfer between the polaritons; thus, a compromise must be made between energetically aligning states and maintaining the charge transfer character.
We note that the cavity field slightly reduces the total charge transfer (see the black dashed line in Fig.~\ref{fig:PNA_ex_dens}).
This is because the charge transfer state also contributes to the other excited states, not just the polaritons.

In Fig.~\ref{fig:PNA_ex_dens_freq} we show how fine-tuning the cavity frequency can be used to change the degree of charge transfer in the lower and upper polaritons. 
This opens another possibility for charge transfer control, with potential for technological applications.

\subsection{\label{sec:pyrrole} Photochemical processes}

We now turn our attention to photochemical processes and the possibility 
of changing the ground state potential energy surface using an 
optical cavity. 
For this purpose, we choose the pyrrole molecule that exhibits conical intersections between the ground state and two low lying excited states.
A detailed analysis of these conical intersections, and of the involved relaxation mechanism, can be found in Ref.~\cite{PicconiPhD2017, PicconiChemPhys2016}.
In $C_{2v}$ symmetry, the ground state is $^1A_1$ while the first two excited states are $^1A_2$ and $^1B_1$. The equilibrium geometry is calculated with CCSD and cc-pVDZ basis set.

\begin{figure}[ht!]
    \centering
    \includegraphics[trim={1.5cm 1cm 1cm 1.5cm},clip,width=8.6cm]{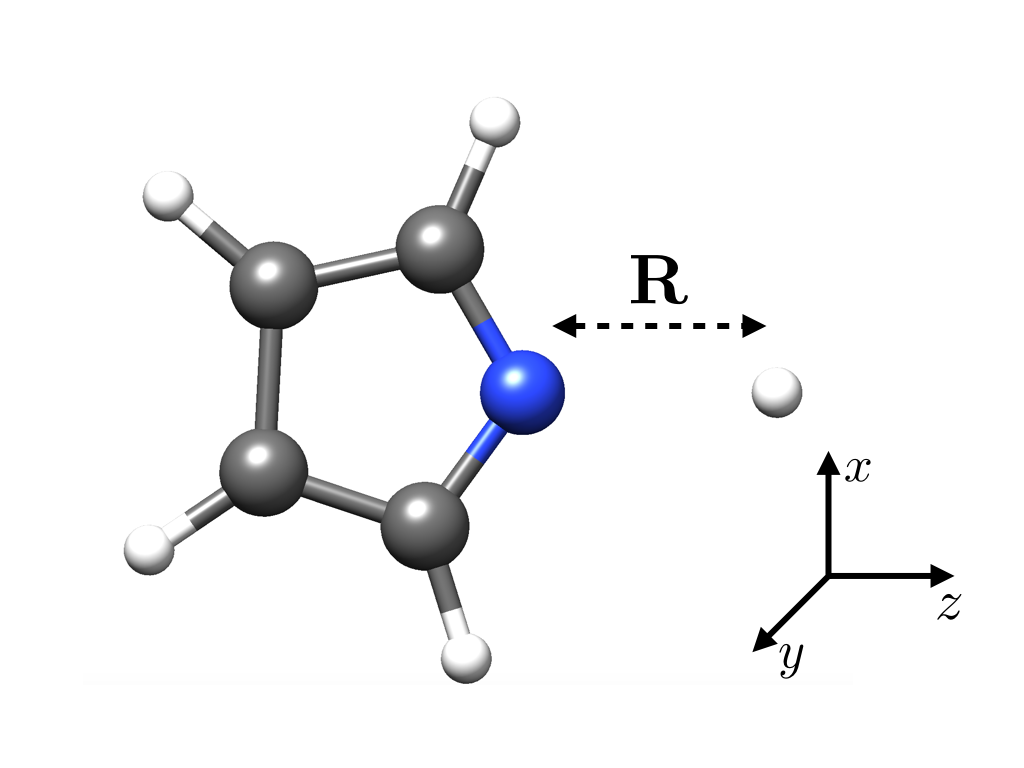}
    \caption{Orientation of pyrrole with indication of the coordinates.}
    \label{fig:pyr_coord_geo}
\end{figure}

We investigate the behavior of the potential energy curves when the NH bond distance $R$ is varied, preserving the $C_{2v}$ symmetry (Fig.~\ref{fig:pyr_coord_geo}).
The polarization of the cavity is chosen as $\vec{e}=\left(\frac{1}{\sqrt{3}},\frac{1}{\sqrt{3}},\frac{1}{\sqrt{3}}\right)$, such that the point group symmetry of the Hamiltonian reduces to $C_1$. The coupling strength is set to $\lambda=0.05$ and the cavity frequency is set in resonance with the $^1B_1$ state at $R=2.0$ \AA  ($\omega_{cav} = 1.06$~eV).

\begin{figure*}[ht!]
    \centering
    \includegraphics[trim={2.5cm 2.0cm 1.5cm 1.4cm},clip,width=0.7\textwidth]{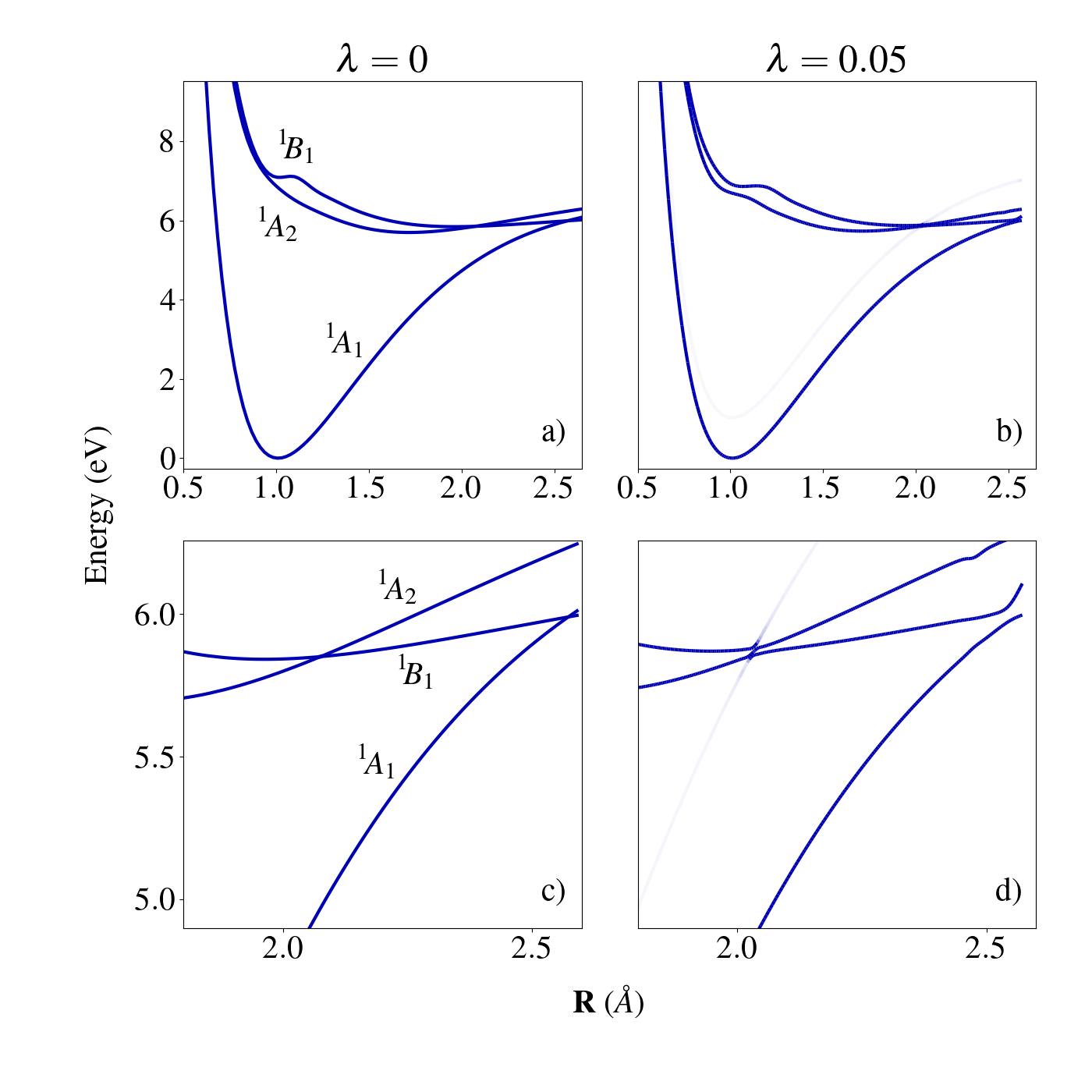}
    \caption{Potential energy curves in QED-CCSD-1 for pyrrole calculated without a),c) and with b),d) the cavity. Panels c) and d) are zooms of a) and b) respectively, in the conical intersections' region. The energies are relative to equilibrium energy. In panel b) and d) the blue colormap indicates the electronic/photonic character of the states.}
    \label{fig:Pyrrole}
\end{figure*}

In Fig.~\ref{fig:Pyrrole}, we show the potential energy curves along the coordinate $R$ with and without the cavity.
Without the cavity ($\lambda = 0$), the CCSD model mostly reproduces the accurate potential energy curves calculated in Refs.~\cite{PicconiPhD2017, PicconiChemPhys2016}.
The main difference is an inverted ordering of the $^1A_2$ and $^1B_1$ states close to the conical intersection.
The correct ordering can be recovered by including triple excitations in the model, as we have confirmed with CC3~\cite{Koch1997} calculations (see Appendix~\ref{app:pyrrole}).
Since the ordering of the states does not change the conclusions, we performed the analysis at the QED-CCSD-1 level.

We first observe the lifting of the degeneracy as shown in Fig.~\ref{fig:Pyrrole}d. In the coupled system, now in $C_1$ symmetry, all states can interact as they have the same symmetry. This unequivocally demonstrates that the cavity can also significantly impact the potential energy surface of the electronic ground state.
This is the first time this phenomenon has been demonstrated in a molecule using an \textit{ab initio} Hamiltonian.
Previously, a similar observation was made with a model Hamiltonian for graphene~\cite{WangPhysRevB2019}.

Interestingly, a coupling strength ($\lambda=0.05$) that produces a relatively small Rabi splitting of $0.03$~eV around 2 {\AA} is able to open a gap almost twice this size, $0.05$ eV, at the conical intersection.
We can rationalize this observation in the following way. The Rabi splitting is mainly due to the bilinear term in the Hamiltonian, whereas the lifting of the degeneracy is mainly due to the self-energy term.
In Appendix~\ref{app:pyrrole} we show that the symmetry breaking is insensitive to basis sets and cavity frequencies.

A consequence of the above observation is the possibility to change relaxation pathways in chemical reactions. The electron-photon coupling may cause intersection seams to move or vanish. Note that the avoided crossing in Fig.~\ref{fig:Pyrrole}d does not exclude the possibility that the seam has been moved to a distorted molecular geometry. In any case, the changes to the energy surfaces may for example lead to reduced relaxation through the conical intersection to the ground state, such that the relaxation can be dominated by radiative processes (which are generally slower). 
Note also that the majority of molecular orientations allow for this type of symmetry breaking. This should make the gap opening experimentally observable in standard temperature conditions where the molecules can rotate freely.

\section{\label{sec:conclusions} Concluding remarks}

We have developed a coupled cluster theory that explicitly incorporates quantized electromagnetic fields, denoted as QED-CC.
This non-perturbative theory can describe molecular photochemistry inside an optical cavity. 
The QED-CC model is a natural extension of the well established coupled cluster model, used in electronic structure theory. 
The method provides a highly accurate description of electron-electron and electron-photon correlation, at least in regions where the electronic ground state is dominated by a single determinant.
These correlations are not accounted for in commonly used model Hamiltonians and mean field methods.
The accuracy is demonstrated by comparison with exact diagonalization (within an orbital basis) in a truncated photon space (QED-FCI). Unlike QED-FCI, the QED-CC hierarchy is computationally feasible also for larger molecules. Extensions of QED-CC to approximately include the environment, such as a solvent, is a natural next step in further developments. However, one must then carefully consider how the quantum field should be incorporated into the environment.

Initially, we investigated the restructuring of potential energy curves in diatomic molecules. 
In particular, we found that the interaction with the cavity creates polaritons which are more bound than the corresponding bare electronic excited states.
Clearly, polaritons have crucial implications on the photochemistry.
For instance, they can alter the relaxation pathways and trap molecules in dark excited states. 
A further study of these phenomena would be very interesting. 

Cavity-induced effects on charge transfer processes were also investigated quantitatively for PNA.
We explained how the cavity restructures the charge inside the molecule, and how these effects could be applied in photovoltaics.

Finally, we demonstrated how the cavity field can be used to manipulate conical intersections in molecules.
We showed that the quantum field is able to lift degeneracies between ground and excited state.
This analysis suggests that new experimental strategies can be developed to manipulate ultrafast molecular relaxation mechanisms through conical intersections.

\section*{Acknowledgements}

We acknowledge Laura Grazioli, Rosario Riso and Christian Sch\"afer for insightful discussions. We acknowledge computing resources through UNINETT Sigma2 - the National Infrastructure for High Performance Computing and Data Storage in Norway, through project number NN2962k. We acknowledge funding from the Marie Sk{\l}odowska-Curie European Training Network “COSINE - COmputational Spectroscopy In Natural sciences and Engineering”, Grant Agreement No. 765739, the Research Council of Norway through FRINATEK projects 263110 and 275506. A.R. was supported by the European Research Council (ERC-2015-AdG694097), the Cluster of Excellence ‘Advanced Imaging of Matter' (AIM), Grupos Consolidados (IT1249-19) and SFB925. The Flatiron Institute is a division of the Simons Foundation.

\section*{Author Contributions}
T.S.H. and E.R. contributed equally to this work.

\vfill
\begin{appendix}

\section{Derivation of QED-HF Theory}\label{app:QED-HF}

Consider the Pauli-Fierz Hamiltonian in the coherent state basis, Eq.~\eqref{eq:sim_trans_H_final}, with a single photon mode,
\begin{eqnarray}\label{eq:app-sim_trans_H_final}
    H &=& H_{e}
    + \omega b^{\dagger}b
    + \frac{1}{2} (\vec{\lambda} \cdot (\vec{d}-\langle\vec{d}\rangle))^2
    \nonumber \\
    &-& \sqrt{\frac{\omega}{2}}(\vec{\lambda} \cdot (\Vec{d}-\langle\Vec{d}\rangle)) (b^{\dagger} + b).
\end{eqnarray}
When averaging over the photon vacuum state $|0\rangle$ and using Eq.~\eqref{eq:identity} we obtain
\begin{widetext}
\begin{eqnarray}\label{eq:app-ham0}
    \langle H \rangle_{0} &=& \sum_{pq} \bigg( h_{pq} + \frac{1}{2} \sum_r (\Vec{\lambda} \cdot \Vec{d}_{pr}
    )(\Vec{\lambda} \cdot \vec{d}_{rq})
    - (\Vec{\lambda}\cdot\langle\Vec{d}\rangle)(\Vec{\lambda} \cdot \Vec{d}_{pq}) \bigg) E_{pq}
    \nonumber\\
    &+& \frac{1}{2} \sum_{pqrs} \Big(
   g_{pqrs}
    + (\Vec{\lambda} \cdot \Vec{d}_{pq}) (\Vec{\lambda}  \cdot \Vec{d}_{rs})
    \Big) e_{pqrs}
    + \frac{1}{2} (\Vec{\lambda} \cdot \langle\Vec{d}\rangle)^2
    + h_{nuc}.
\end{eqnarray}
\end{widetext}
This operator has the same form as the electronic Hamiltonian in Eq.~\eqref{eq:H_e}, with modified integrals and constants. These modified integrals can be inserted directly into the expression for the inactive electronic Fock matrix,
\begin{equation}
    F^e_{pq} = h_{pq} + \sum_i (2g_{pqii} - g_{piiq}),
\end{equation}
and we obtain Eq.~\eqref{eq:fock-matrix} for a single mode.
This procedure is easily generalized to the multimode case.

We now consider the origin invariance in QED-HF by shifting the dipole by a constant, $\Vec{d} \rightarrow \Vec{d} + \Delta \Vec{d}$.
For neutral molecules the dipole moment operator is origin invariant, $\Delta \Vec{d}=0$, but for charged molecules this is not the case.
Nevertheless, since the Hamiltonian~\eqref{eq:app-sim_trans_H_final} is invariant, the energy is also invariant. On the other hand, the inactive Fock matrix in Eq.~\eqref{eq:fock-matrix} is not invariant. As seen from the shifted Fock matrix,
\begin{widetext}
\begin{equation}
\begin{pmatrix}
    F_{ij} & F_{ib} \\
    F_{aj} & F_{ab}
\end{pmatrix}
\rightarrow
\begin{pmatrix}
    F_{ij} - (\Vec{\lambda} \cdot \Delta \Vec{d})(\Vec{\lambda} \cdot \Vec{d}_{ij}) - \frac{1}{2} (\Vec{\lambda} \cdot \Delta \Vec{d})^2 \delta_{ij}
   & F_{ib} \\
    F_{aj} &
    F_{ab} + (\Vec{\lambda} \cdot \Delta \Vec{d})(\Vec{\lambda} \cdot \Vec{d}_{ab}) + \frac{1}{2} (\Vec{\lambda} \cdot \Delta \Vec{d})^2 \delta_{ab}
\end{pmatrix}
,
\end{equation}
\end{widetext}
the occupied-virtual block of the Fock matrix, $F_{ib}$ and $F_{aj}$, are unchanged, whereas the purely occupied and virtual blocks, $F_{ij}$ and $F_{ab}$, are dependent on the origin. Explicitly, for the occupied-virtual block of the Fock matrix we have
\begin{eqnarray}
    F_{jb} \rightarrow F_{jb}
    &+& \frac{1}{2} \sum_{a} 
    \delta_{ab}
    (\Vec{\lambda}\cdot\Delta\Vec{d})
    (\Vec{\lambda}\cdot\Vec{d}_{ja})
    \nonumber \\
    &-& \frac{1}{2} \sum_{i}
    \delta_{ij}
    (\Vec{\lambda}\cdot\Delta\Vec{d})
    (\Vec{\lambda}\cdot\Vec{d}_{ib})
    = F_{jb}.
\end{eqnarray}

\section{The QED-CCSD-1 ground state equations}\label{app:QED-CC-SD-S-DT}

Using the QED-CCSD-1 cluster operator defined in Eq.~\eqref{eq:QED_CCSD_T}, the coupled cluster ground state equations, see Eqs.~\eqref{eq:CCsolveq_1} and \eqref{eq:CCsolveq_2}, take the form
\begin{align}
    \begin{split}\label{eq:CCSDsolvq_1}
        \langle R \rvert \bar{H} \lvert R \rangle =
        \langle R \rvert
        \mathcal{H} + [\mathcal{H},T_2] +
        [\mathcal{H}, S^1_1] +
        [\mathcal{H}, \Gamma^1]
        \lvert R\rangle\\
    \end{split}\\
    &\nonumber\\
   \begin{split}\label{eq:CCSDsolvq_2}
        \langle \mu, 0\rvert
        \bar{H} \lvert R \rangle
        &=
        \langle \mu,0 \rvert
        \mathcal{H} + [\mathcal{H},T_2]
        + \frac{1}{2} [[\mathcal{H},T_2],T_2]\\
        &+ [\mathcal{H},S^1_1]
        + [[\mathcal{H},S^1_1],T_2]
        + [\mathcal{H},S^1_2]\\
        &+ [\mathcal{H},\Gamma^1]
        + [[\mathcal{H},\Gamma^1 ],T_2]
        \lvert R\rangle\\
   \end{split}\\
   &\nonumber\\
   \begin{split}\label{eq:CCSDsolvq_3}
        \langle \mu,1 \rvert \bar{H}\lvert R \rangle &=
        \langle \mu,1 \rvert \mathcal{H} + [\mathcal{H},T_2] + [[\mathcal{H},S^1_2 ],T_2] +
        [\mathcal{H},S^1_1] \\
        &+ [\mathcal{H},S^1_2] + 
        \frac{1}{2} [[\mathcal{H},S^1_1],S^1_1] +
        [[\mathcal{H},S^1_1],S^1_2] \\
        &+ [\mathcal{H}, \Gamma^1] + 
        [[\mathcal{H}, \Gamma^1], T_2] +
        [[\mathcal{H},\Gamma^1],S^1_1]\\
        &+ [[\mathcal{H},\Gamma^1],S^1_2] +
        [[\mathcal{H},S^1_1],T_2] \lvert R\rangle\\
   \end{split}\\
   &\nonumber\\
   \begin{split}\label{eq:CCSDsolvq_4}
        \langle \text{HF},1\rvert \bar{H} \lvert R \rangle =
        \langle \text{HF} ,1 \rvert \mathcal{H} + [\mathcal{H},\Gamma^1] + [\mathcal{H},S^1_1] +
        [\mathcal{H},S^1_2] \lvert R\rangle.\\
   \end{split}
\end{align}
We have here introduced the notation,
\begin{equation}\label{eq:app-t1-ham}
    \mathcal{H} = \exp(-T_1) H \exp(T_1).
\end{equation}
This operator can be expressed as $H$ with transformed one- and two-electron integrals~\cite{HelgakerBook2000}.

\begin{figure*}[ht!]
    \centering
    \includegraphics[trim={0.5cm 2.8cm 0cm 6cm},clip,width=\textwidth]{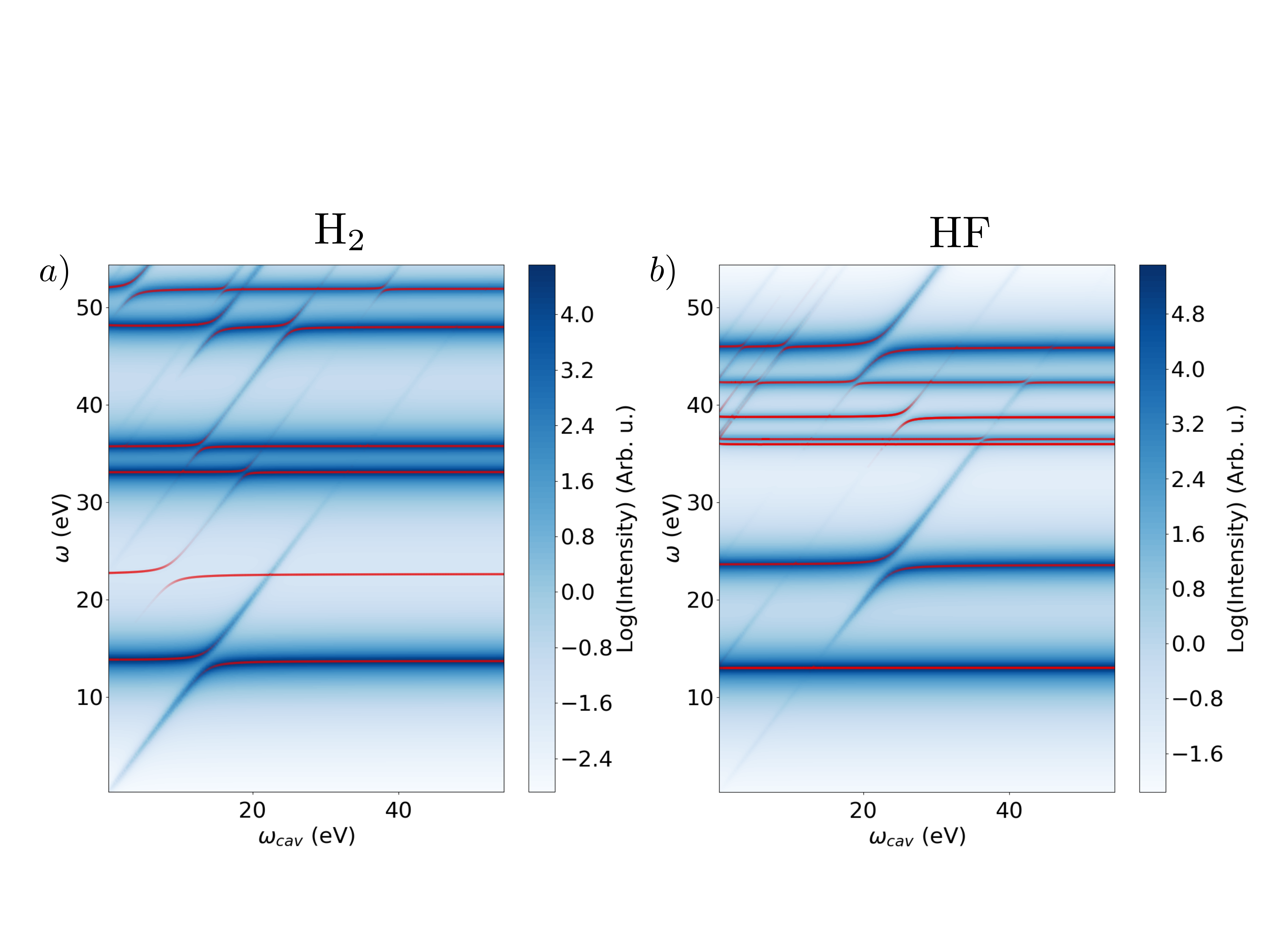}
    \caption{Comparison between QED-CCSD-1 (red solid lines) and QED-FCI (countour map in background) energy dispersion plots for a) H$_2$ and b) HF. The intensity of the red lines indicates the electronic/photonic  contribution to the states. $\lambda = 0.05$ and $\eta = 0.01$ a.u.}
    \label{fig:confFCICC}
\end{figure*}
\section{The QED-CCSD-1 excited state equations}\label{app:EOM-QED-CC-SD-S-DT}

The expressions for the Jacobian in Eq.~\eqref{eq:qed-jacobian-matrix} are derived using the commutator expansion. We obtain
\begin{align}
    \begin{split}
        \langle \mu, 0 \rvert [\bar{H}, \tau_\nu] \lvert R \rangle &=
        \langle \mu, 0 \rvert
        \left[ \mathcal{H}
        + [\mathcal{H}, T_2]
        + [\mathcal{H}, S^1_1]\right.\\
        &+ \left.[\mathcal{H}, S^1_2]
        + [\mathcal{H}, \Gamma_1]
        , \tau_\nu \right] \lvert R \rangle\\
    \end{split}\\
    &\nonumber\\
    \begin{split}
        \langle \mu, 0 \rvert [ \bar{H}, b^{\dagger}]  \lvert R \rangle &=
        \langle \mu, 0 \rvert
        \left[ \mathcal{H}
        + [\mathcal{H}, T_2]
        , b^{\dagger} \right] \lvert R \rangle\\
    \end{split}\\
    &\nonumber\\
    \begin{split}
        \langle \mu, 0 \rvert [ \bar{H}, \tau_\nu b^{\dagger}] \lvert R \rangle &=
        \langle \mu, 0 \rvert
        \left[ \mathcal{H}
        + [\mathcal{H}, T_2],
        \tau_\nu b^{\dagger} \right] \lvert R \rangle\\
    \end{split}\\
    &\nonumber\\
    \begin{split}
        \langle \text{HF}, 1 \rvert [\bar{H}, \tau_\nu] \lvert R \rangle &=
        \langle \text{HF}, 1 \rvert
        \left[ \mathcal{H}
        + [\mathcal{H}, S^1_1]\right.\\
        &+ \left.[\mathcal{H}, S^1_2]
        + [\mathcal{H}, \Gamma^1]
        , \tau_\nu \right] \lvert R \rangle\\
    \end{split}\\
    &\nonumber\\
    \begin{split}
        \langle \text{HF}, 1 \rvert [\bar{H}, b^{\dagger}] \lvert R \rangle &=
        \langle \text{HF}, 1 \rvert
        \left[ \mathcal{H}    
        + [\mathcal{H}, S^1_1]
        , b^{\dagger}\right] \lvert R \rangle\\
    \end{split}\\
    &\nonumber\\
    \begin{split}
        \langle \text{HF}, 1 \rvert [\bar{H}, \tau_\nu b^{\dagger}] \lvert R \rangle &=
        \langle \text{HF}, 1 \rvert
        \left[ \mathcal{H}
        ,\tau_\nu b^{\dagger}\right] \lvert R \rangle\\
    \end{split}\\
    &\nonumber\\
    \begin{split}
        \langle \mu, 1 \rvert [\bar{H}, \tau_\nu] \lvert R \rangle &=
        \langle \mu, 1 \rvert
        \big[ \mathcal{H} 
        + [\mathcal{H}, T_2]
        + [\mathcal{H}, S^1_1]\\
        &+ [\mathcal{H}, S^1_2]
        + [\mathcal{H}, \Gamma^1]\\
        &+ [[\mathcal{H}, \Gamma^1], S^1_1]
        + [[\mathcal{H}, \Gamma^1], S^1_2]
        , \tau_\nu \big] \lvert R \rangle\\
    \end{split}\\
    &\nonumber\\
    \begin{split}
        \langle \mu, 1 \rvert [\bar{H}, b^{\dagger}] \lvert R\rangle &=
        \langle \mu, 1 \rvert
        \left[ [\mathcal{H}, S^1_1]
        + [\mathcal{H}, S^1_2]
        , b^{\dagger}\right] \lvert R\rangle\\
    \end{split}\\
    &\nonumber\\
    \begin{split}
        \langle \mu, 1 \rvert [\bar{H}, \tau_\nu b^{\dagger}]             
        \lvert R \rangle &=
        \langle \mu, 1 \rvert
        \left[ \mathcal{H}
        + [\mathcal{H}, T_2]
        + [\mathcal{H}, S^1_1]\right.\\
        &+ \left.[\mathcal{H}, S^1_2]
        + [\mathcal{H}, \Gamma^1]
        , \tau_\nu b^{\dagger} \right] \lvert R \rangle\\
    \end{split}
\end{align}
For completeness, we also give the expressions for the $\eta_{\nu}$ block of Eq.~\eqref{eq:eomH}:
\begin{align}
& \langle R \rvert \bar{H} \lvert \nu, 0 \rangle = \langle R\rvert
\mathcal{H} + [\mathcal{H}, T_2] + [\mathcal{H}, S^1_1]
\lvert \nu, 1 \rangle
\\
&\nonumber\\
& \langle R \rvert \bar{H} \lvert \text{HF}, 1 \rangle = \langle R \rvert
\mathcal{H}
\lvert \text{HF}, 1 \rangle
\\
&\nonumber\\
& \langle R \rvert \bar{H} \lvert \nu, 1 \rangle = \langle R \rvert
\mathcal{H}
\lvert \nu, 1 \rangle.
\end{align}

\section{Electronic weights} \label{app:fractions}

The ground state electronic weights $w^e_{gs}$ in QED-CCSD-1 are calculated by projecting the ground state wave function $|\text{CC} \rangle$ on the elementary electronic basis, that is,
\begin{eqnarray}\label{eq:app-fraction-gs}
    w^e_{gs} = \sqrt{\frac{\langle \text{CC} | P_{el} | \text{CC} \rangle}{\langle \text{CC} | P | \text{CC} \rangle} }.
\end{eqnarray}
The electronic and total projection operators are here defined as 
\begin{eqnarray}
    P_{el} &=& | R \rangle \langle R | + \sum_\mu  | \mu,0 \rangle \langle \mu,0 |,
    \\
    P &=& | R \rangle \langle R | + | \text{HF},1 \rangle \langle \text{HF},1 | \nonumber\\
    &+& \sum_\mu \left( | \mu,0 \rangle \langle \mu,0 | + | \mu,1 \rangle \langle \mu,1 | \right).
\end{eqnarray}
Excited states weights, $w^e_{es}$, are calculated in an approximate way using the norm of the electronic part of the excitation vector $\vec{\mathcal{R}}$,
\begin{equation}\label{eq:app-fraction-es}
    w^e_{es}=
    \sqrt{\frac{\sum_{\mu} (\mathcal{R}_{\mu,0})^2 }{||\vec{\mathcal{R}}||^2}}.
\end{equation}
In principle an equation equivalent to Eq.~\eqref{eq:app-fraction-gs} should be used also in this case, substituting $|\text{CC} \rangle$ with $|\mathcal{R} \rangle$. Considering that we are only interested in a qualitative estimate of the weights, the approximate form Eq.~\eqref{eq:app-fraction-es} is adequate.

\section{Comparison of QED-CCSD-1 and QED-FCI}\label{app:comp_CC_FCI}

We here compare QED-CCSD-1 and QED-FCI. For the latter method, we performed an exact diagonalization of the Hamiltonian in Eq.~\eqref{eq:QEDHam} with one photonic excitation, in order to obtain a consistent comparison.

We used a 3-21G basis~\cite{BinkleyJAmChemSoc1980} for H$_2$ and a STO-3G~\cite{HehreJChemPhys1969} basis for HF, with internuclear distances $R_{\text{H}_2}=1.0$ {\AA} and $R_{\text{HF}}=0.917$ {\AA}.
In both cases, the coupling value $\lambda = 0.05$ is used. 

In Fig.~\ref{fig:confFCICC}, we show the energy dispersion with respect to the cavity frequency $\omega_{cav}$.
In this figure we use the QED-FCI linear response spectral function,
\begin{equation}
    A(\omega) = -\frac{1}{\pi} \mathrm{Im} \sum_{n\neq0}\frac{\langle\Psi_0|\sum_{ij}a_i^{\dagger}a_j|\Psi_n\rangle\langle\Psi_n|\sum_{ij}a_i^{\dagger}a_j|\Psi_0\rangle}{\omega - (E_n - E_0) + i\eta},
\end{equation}
where $|\Psi_n\rangle$ are the eigenfunctions of the QED Hamiltonian.
In this case, we calculated the density-density spectral function instead of the more appropriate optical spectrum (with transition dipole moments) in order to compare the states independently from the selection rules. 
Coupled cluster results are displayed in Fig.~\ref{fig:confFCICC} as red lines with electronic weights.

\begin{figure}[ht!]
    \centering
    \includegraphics[trim={1cm 1cm 1cm 1cm},clip,width=8.6cm]{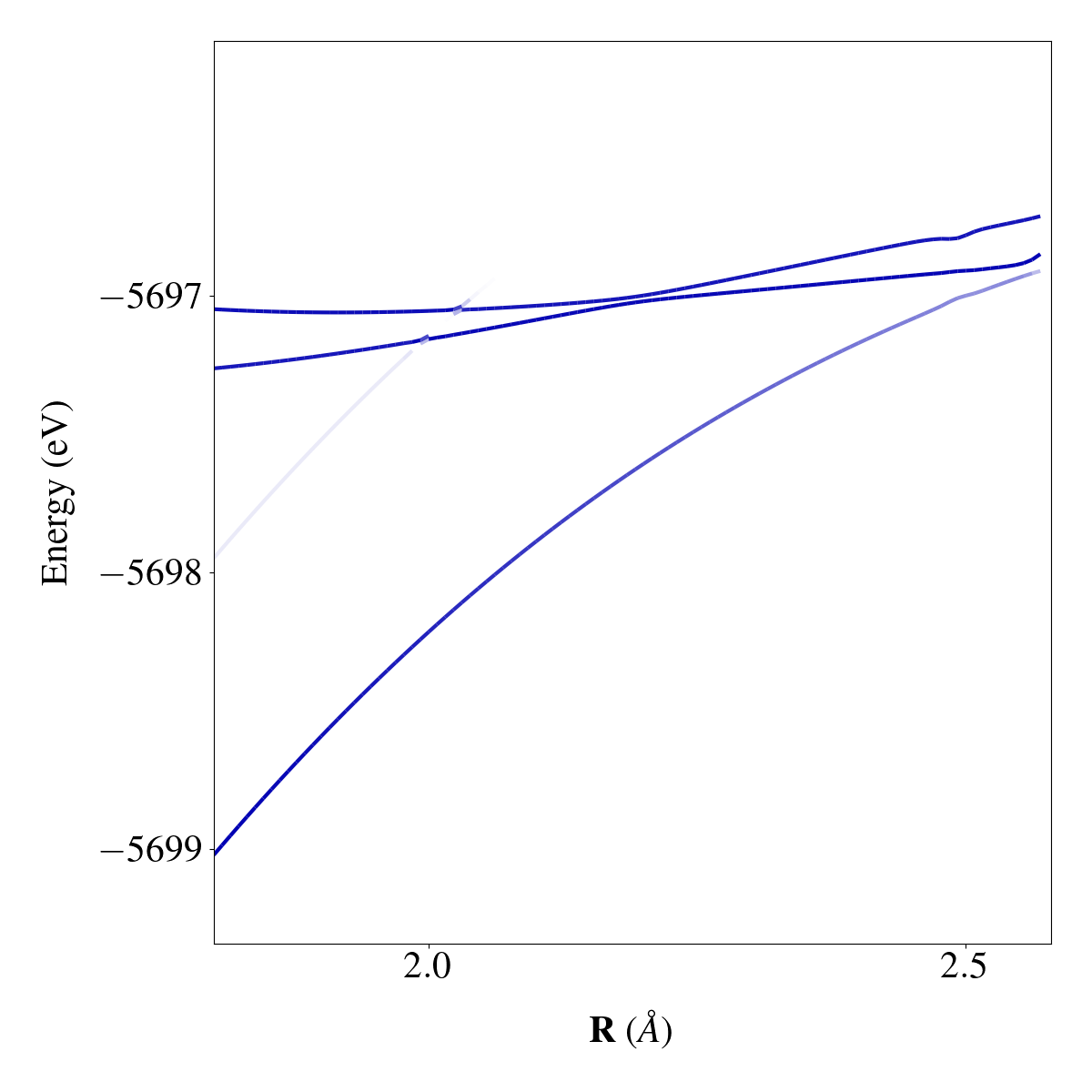}
    \caption{Potential energy curves in QED-CCSD-1 for pyrrole calculated with cavity for $\lambda=0.05$ using the aug-cc-pVDZ basis set. The blue colormap indicates the electronic/photonic character of the states.}
    \label{fig:pyr_aug}
\end{figure}

For both molecules we observe an excellent agreement.
The only noticeable difference is the absence of a few electronically excited states in the FCI spectrum.
These states have large double excitation character, and thus, small contributions to the spectral function. 
Notice here that QED-CCSD-1 is very accurate also for HF, where the electronic structure is not exact in CCSD.

\section{Additional results for pyrrole}\label{app:pyrrole}

\begin{figure}[ht!]
    \centering
    \includegraphics[trim={1cm 1cm 1cm 1cm},clip,width=8.6cm]{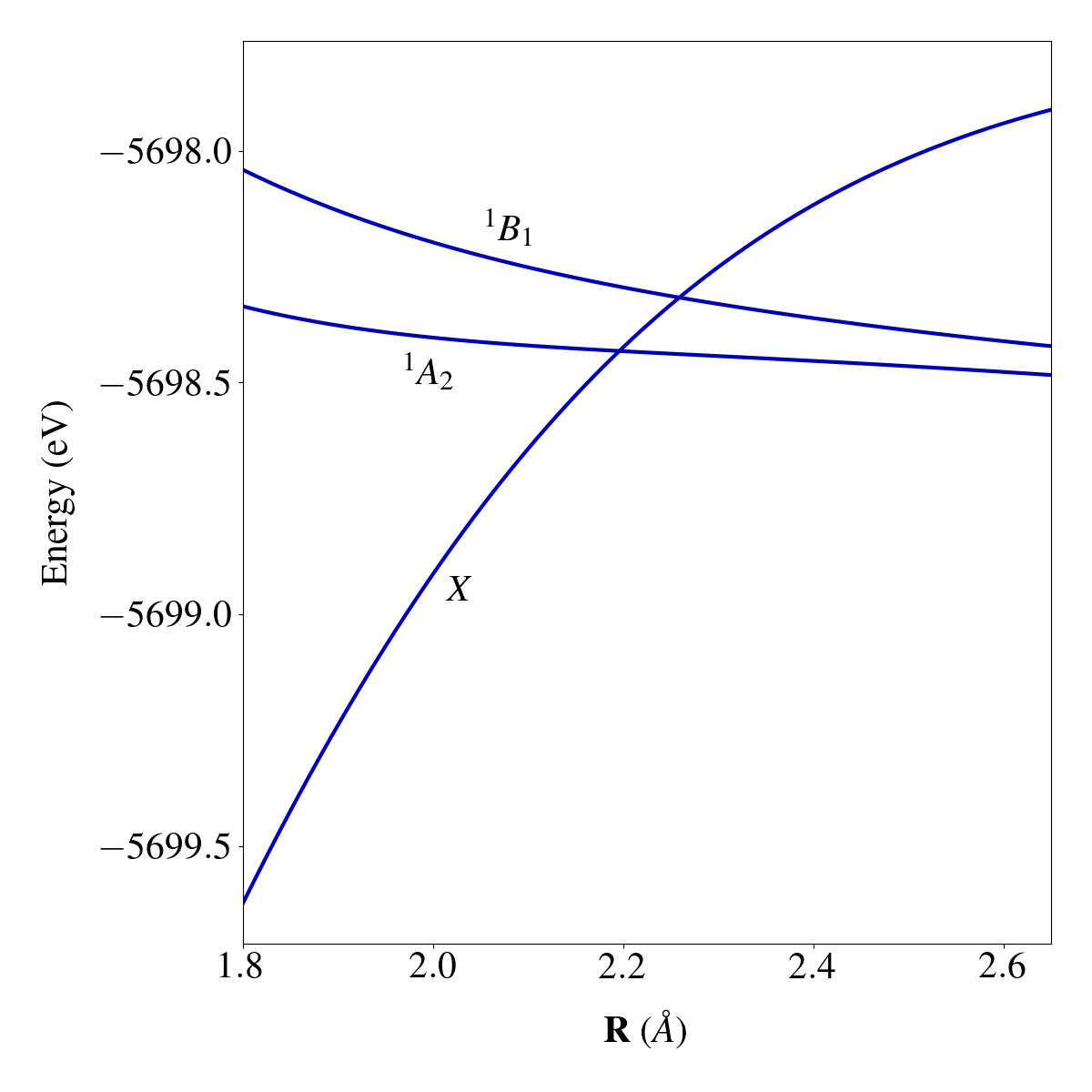}
    \caption{Potential energy curves for pyrrole without the cavity at the CC3 level of theory with cc-pVDZ.}
    \label{fig:pyr_cc3}
\end{figure}
We also provide some additional results that are important to validate the accuracy of QED-CCSD-1 for pyrrole.
Firstly, we address the basis set appropriateness, as electromagnetic fields may require more diffuse basis functions. 
In Fig.~\ref{fig:pyr_aug} potential energy curves calculated using an aug-cc-pVDZ~\cite{DunningJChemPhys1989} basis are shown. 
The inclusion of diffuse functions does not change the general qualitative picture described in Section~\ref{sec:pyrrole}, confirming the accuracy of our predictions.
In particular, the position of the intersections is nearly unaffected by the size of the basis and the qualitative shape of the potential energy curves is unchanged.
\begin{figure}[ht!]
    \centering
    \includegraphics[trim={1cm 1cm 1cm 1cm},clip,width=8.6cm]{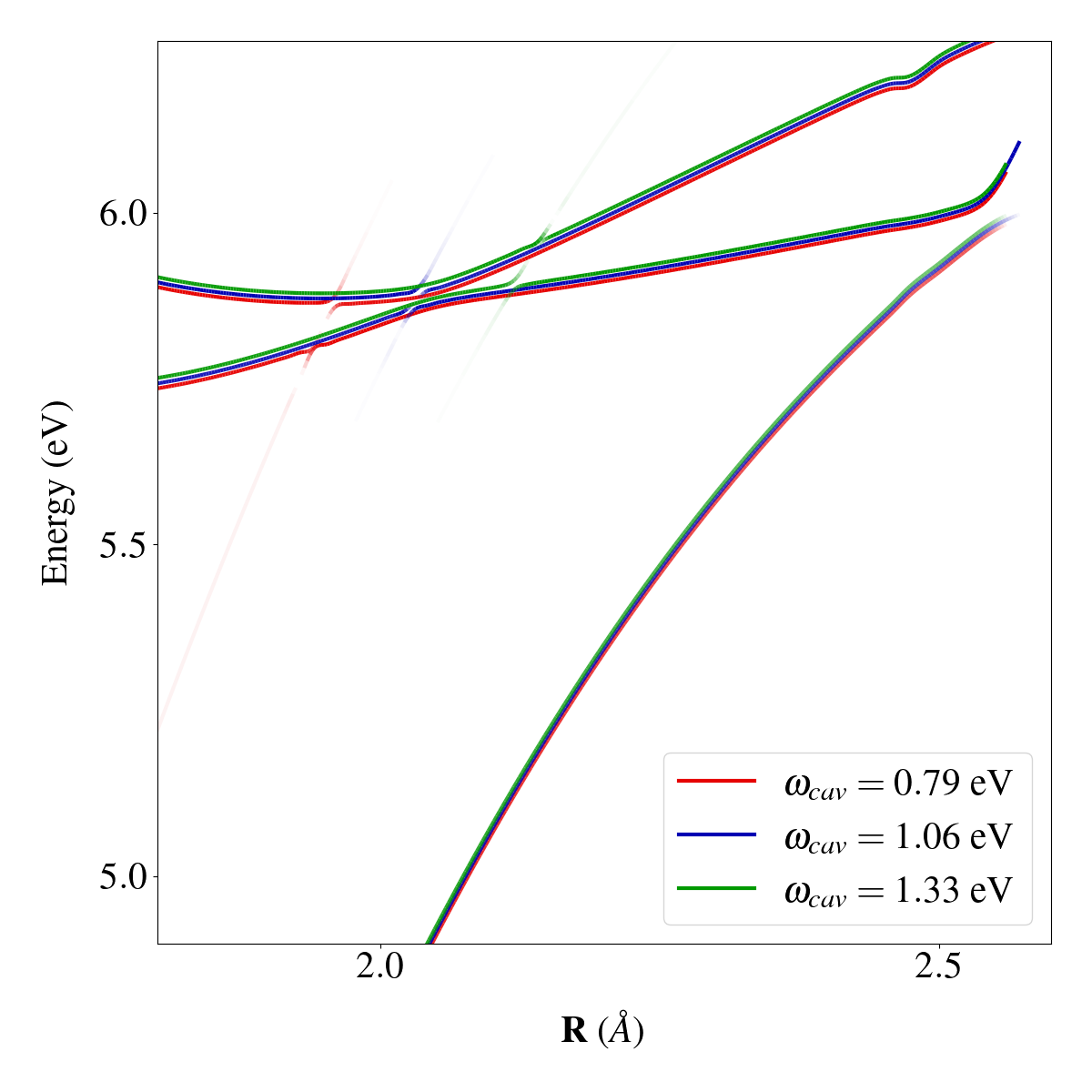}
    \caption{Potential energy curves in QED-CCSD-1 for pyrrole with cavities at different frequencies $\omega_{cav}$, coupling strength $\lambda=0.05$ and with cc-pVDZ. The energies are relative to equilibrium energy.}
    \label{fig:pyr_freq}
\end{figure}

In Section~\ref{sec:pyrrole}, we noted that CCSD gives and incorrect ordering of the $^1A_2$ and $^1B_1$ excited states.
This can be rectified by including triple excitations in the electronic treatment, as shown with CC3 in Fig.~\ref{fig:pyr_cc3}.

The opening of the conical intersection is quite robust with respect to changes in the cavity frequency, $\omega_{cav}$,
as seen in Fig.~\ref{fig:pyr_freq}.
This supports the claim that the dipole self-energy term is mainly responsible for lifting the degeneracy.
The position of the Rabi splitting is, as expected, highly sensitive to the cavity frequency.

\end{appendix}

\clearpage
\bibliography{bib}

\clearpage
\section*{TOC}
\includegraphics[width=8.6cm]{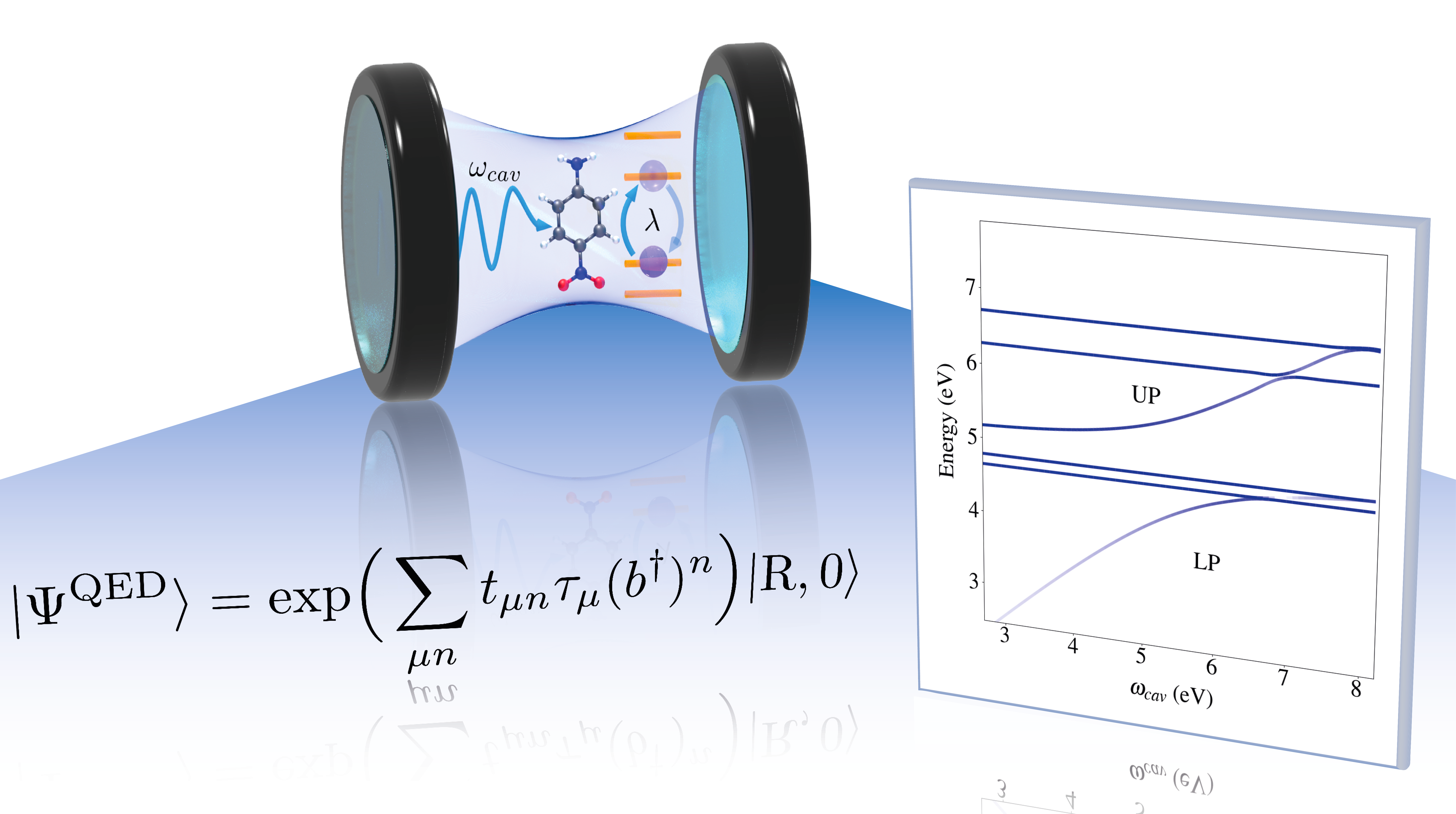}

\end{document}